\documentclass[12pt]{article}
\usepackage{amssymb}

\makeatletter

\usepackage{cite}

\usepackage{latexsym}
\usepackage{amsfonts}

\usepackage{amsmath}

\usepackage[subnum]{cases}

\@addtoreset{equation}{section}

\newcommand{\eref}[1]{(\ref{#1})}

\def\be{\begin{equation}}
\def\ee{\end{equation}}
\def\ba{\begin{eqnarray}}
\def\ea{\end{eqnarray}}
\def\pa{\partial}

\def\nn{\nonumber}
\def\ve{\varepsilon}

\def\a{\alpha}
\def\bt{\beta}
\def\g{\gamma}
\def\G{\Gamma}
\def\D{\Delta}

\def\La{\Lambda}
\def\dl{\delta}
\def\la{\lambda}

\def\s{\sigma}

\def\m{\mu}
\def\n{\nu}
\def\ra{\rightarrow}
\def\vp{\varphi}

\def\w{\widetilde}

\def\dis{\displaystyle}
\def\wh{\widehat}

\long\def\symbolfootnote[#1]#2{\begingroup%
\def\thefootnote{\fnsymbol{footnote}}\footnote[#1]{#2}\endgroup}

\makeatother

\begin{document}
\begin{titlepage}

\global\long\def\LimV{\underset{V\rightarrow1}{\mathrm{Lim}}\;}
\global\long\def\LimE{\underset{\varepsilon\rightarrow0}{\mathrm{Lim}}\;}

\begin{flushright}
Preprint DFPD/2016/TH05\\
April 2016\\

\par\end{flushright}

\vspace{0truecm}

\begin{center}
\textbf{\Large Dynamics of self-interacting strings and energy-momentum conservation} \vskip0.3truecm
\par\end{center}{\Large \par}

\vspace{0.5cm}

\begin{center}
Kurt Lechner\symbolfootnote[2]{kurt.lechner@pd.infn.it}
\par\end{center}

\vskip1truecm

\begin{center}
\textit{Dipartimento di Fisica e Astronomia, Universit\`a degli Studi di Padova, Italy}
\par
\end{center}

\begin{center}
\textit{and}
\par\end{center}

\begin{center}
\textit{
 INFN, Sezione di Padova,}
\par\end{center}

\begin{center}
\textit{Via F. Marzolo, 8, 35131 Padova, Italy}
\par\end{center}

\vspace{0.3cm}

\begin{abstract}
\vskip0.2truecm

Classical strings coupled to a metric, a dilaton and an axion, as conceived by superstring theory, suffer from ultraviolet divergences due to self-interactions. Consequently, as in the case of radiating charged  particles, the corresponding effective string dynamics can not be derived from an action principle. We propose a {\it fundamental principle} to build this dynamics, based on local energy-momentum conservation in terms of a well-defined  distribution-valued  energy-momentum tensor. Its continuity equation implies a finite equation of motion for self-interacting strings. The construction is carried out explicitly for strings in uniform motion in arbitrary space-time dimensions, where we establish cancelations of ultraviolet divergences which parallel superstring non-renormalization theorems. The uniqueness properties of the resulting dynamics are analyzed.

\vspace{0.1cm}

\end{abstract}
\end{titlepage}

\newpage

\baselineskip 6mm

\tableofcontents

\section{Introduction}

In the same way as charged particles in four space-time dimensions are subject to divergent electromagnetic self-interactions, generic  charged  extended objects, $p$-branes, in $D$ space-time dimensions are subject to infinite self-interactions. The reason for this is that the fields created by a brane become singular on the brane world-volume, meaning that the {\it self-fields}, and hence the {\it self-forces}, are infinite.
A - in a certain sense dramatic - consequence of these ultraviolet divergences is that the theory of self-interacting branes can not be derived from a variational principle: while the original fundamental equations of motion for fields and branes  follow of course from an action principle, once one substitutes the fields resolving the formers in the equations of motion of the latter, the resulting equations are divergent. If one isolates and subtracts - adapting whatever prescription - the infinities, the resulting non-local equations of motion of the brane do no longer follow from an action principle. This in turn implies that the conservation laws, in particular energy-momentum conservation, can not be derived from N\"other's theorem, see {\it e.g.} \cite{R,LM,KL0} for the case of self-interacting charged particles and dyons in $D=4$. Within this approach one looses thus the control over energy-momentum conservation.

More precisely ultraviolet divergences show up in brane theory in two, a priori, unrelated physical quantities: $i)$ in the {\it self-force} of the brane, {\it i.e.} the force exerted by the field generated by the brane on the brane itself, as explained above, and $ii)$ in the $D$-momentum contained in a volume $V$ enclosing (a portion of) the brane. Although the origins of the divergences appearing in these two quantities - the self-force and the $D$-momentum - are the same, {\it i.e.} the bad ultraviolet behavior of the field in the vicinity of the brane, their cures require actually two distinct unrelated procedures \cite{KL1}.

To cure the divergent self-force one may proceed, as anticipated above, regularizing the field produced by the brane in some way, evaluating it on the brane and trying then to isolate and subtract the divergent terms.

The cure of the infinite $D$-momentum requires instead the construction of a well-defined {\it distribution-valued} energy-momentum tensor and offers - at the same time - a strategy for the derivation of the self-force, that is alternative to the approach described above and overcomes its main drawback, {\it i.e.} the missing control over energy-momentum conservation.
It works as follows.

Generically the standard total energy-momentum tensor has the structure
\be
\tau^{\m\n}=\tau^{\m\n}_{\rm field}+ \tau_{\rm kin}^{\m\n}, \quad\quad \tau_{\rm kin}^{\m\n}=M\int l^{\m\n}\dl^D(x-y(\s))\sqrt{\g}\,d^2\s,
\ee
where $\tau_{\rm kin}^{\m\n}$  is the free {\it kinetic} energy-momentum tensor of the brane (with $M$ the brane tension and $y^\m(\s)$ the brane coordinates, see sections \ref{aad} and \ref{eom} for the notations) and $\tau^{\m\n}_{\rm field}$ is the {\it bare} energy-momentum tensor produced by the fields\footnote{Actually in a generic brane- or string-model, as the one considered in this paper, this tensor is given by a sum $\tau^{\m\n}_{\rm field} = \tau^{\m\n}_f+ \tau^{\m\n}_{int}$, where $\tau^{\m\n}_f$ depends only on the fields and is supported on the bulk, and $\tau^{\m\n}_{int}$ is a field-brane interaction-term supported on the world-volume.}: while the fields - solutions of linear d'Alembert equations - are by definition distributions, the tensor $\tau^{\m\n}_{\rm field}$ - a product of the fields - is {\it not} a distribution. Consequently, $i)$ the $D$-momentum of the field
\[
P^\m_V=\int_V  \tau^{0\m}_{\rm field}\, d^3x
\]
contained in a volume $V$  is in general divergent and, $ii)$ it makes no sense to evaluate the divergence $\pa_\m \tau^{\m\n}_{\rm field}$ to analyze the conservation properties of $\tau^{\m\n}$. The cure of these pathologies requires the construction of a {\it renormalized} distribution-valued energy-momentum tensor $T^{\m\n}_{\rm field}$, out of $\tau^{\m\n}_{\rm field}$. A - in principle standard - way to do this consists in the introduction of a regularization - preserving possibly Lorentz- as well as reparameterization-invariance - and the subsequent subtraction from the regularized energy-momentum tensor $(\tau^{\m\n}_{\rm field})_{reg}$  of {\it divergent  local  counterterms}, {\it i.e.} of counterterms supported on the brane that do not converge to distributions as the regularization is removed. By construction the resulting energy-momentum tensor $T^{\m\n}_{\rm field}$ is a distribution and admits hence a well-defined divergence, supported on the word-volume,
\be\label{prel0}
 \pa_\m T^{\m\n}_{\rm field}  =-\int{\cal S}^\n\dl^D(x-y(\s))\sqrt{\g} \,d^p\s,
\ee
where the vector ${\cal S}^\n$ is going to become the {\it finite} self-force of the brane. In fact, for the divergence of  the renormalized total energy-momentum tensor $T^{\m\n}=T^{\m\n}_{\rm field}+ \tau_{\rm kin}^{\m\n}$ one obtains now
\be\label{prel}
\pa_\m T^{\m\n}  =\int \left(M\D_iU^{\n i}- {\cal S}^\n\right)\dl^D(x-y(\s))\sqrt{\g}\,d^p\s,
\ee
where the quantity $\D_iU^{\n i}$ represents the generalized acceleration of the brane. Upon requiring local energy-momentum conservation one derives then the equation of motion for the brane coordinates
\be\label{fam}
M \D_iU^{\n i}= {\cal S}^\n.
\ee

This strategy to derive the self-force may however encounter an obstacle: it can happen that the vector ${\cal S}^\n$ in \eref{prel0} is not a pure {\it multiplication} operator but contains also terms involving derivatives acting on the $\delta$-function, as for example  ${\cal S}^\n \sim \pa^\n$. In this case there would be no equation of motion for the brane ensuring the vanishing of $\pa_\m T^{\m\n}$. This obstacle can be faced through the {\it  finite-counterterm ambiguity} inherent in any renormalization process in physics - in the present case the fact that after the subtraction  of {\it divergent} local counterterms, the renormalized energy-momentum tensor is defined only modulo {\it finite} local counterterms.

The general strategy just described has been envisaged in \cite{KL1}, where a $p$-brane interacting minimally with a $(p+1)$-form potential in $D$ dimensions has been considered, based on previous work facing the analogous problem for massive \cite{LM} as well as massless \cite{AL1,AL2,KL2} point-charges in four dimensions. The present paper represents the first step in the application of this method to the physically more interesting case of the low energy effective
superstring theory, compactified to dimensions $D<10$, where the string  couples to the metric $g_{\m\n}$, the dilaton $\Phi$ and the axion field $B_{\m\n}$. Particular attention will be paid to four-dimensional space-time. We will actually consider two prototype models: a) the {\it general model}, where a certain set of free parameters, or coupling constants, assume generic values, and b) the {\it fundamental string model}, where these parameters are tied by the special relations \eref{fundpar} predicted by ten-dimensional superstring theory.

The problem of ultraviolet divergences and self-interactions of strings moving in a space-time of dimension $D\ge4$  has a long history, especially w.r.t. the problem of tension renormalization and the related finiteness/divergence properties of the self-force and the self-energy. A far from exhaustive literature with this respect is \cite{DH,CHH,DQ,DGHRR,BS,BC1,C,BC2,BD12,BD12bis,CBU,BCM}; for some recent results on the same problem for point-particles see {\it e.g.} \cite{FZ1,FZ2,HFT}. As observed above, by-hand subtractions of divergences from the self-force or from the self-energy - as the ones performed in these references - in general do not ensure energy-momentum conservation. On the contrary the core of our approach is a systematic renormalization of the energy-momentum tensor, comprising $i)$ a covariant separation of the   - in the sense of distributions - {\it divergent counterterms}, $ii)$ the identification of possible finite counterterms and eventually, $iii)$  the implementation of energy-momentum conservation and the consequent derivation of
the self-force. In the present paper the implementation of this program will be carried  out  explicitly for {\it flat} strings, {\it i.e.} for strings in uniform motion, already a
non-trivial task, although in this case the self-force is expected to vanish. Being based essentially on the criteria of {\it finiteness} and {\it energy-momentum conservation}, we regard our approach as a {\it fundamental principle} for the determination of the dynamics of self-interacting extended objects. A particularly powerful aspect of the method - that supports its universality  further - is that it is able to control even strong ultraviolet singularities, as for example the violent divergences generated by growing space-time dimensions or the {\it a priori} uncontrollable divergences generated by objects moving at the speed of light \cite{KL2}.

Since with this respect the contribution of the gravitational self-energy is of fundamental importance, we have to face the problem of which gravitational energy-momentum {\it pseudo-tensor}, and hence which {\it total} energy-momentum pseudo-tensor, we choose. To test the ``stability'' of our construction against different choices we resort to three {\it frameworks}: a) in the, in a certain sense {\it hybrid}, {\it Dirac} framework the gravitational pseudo-tensor  \cite{PAM} is based on the N\"other procedure,  while the matter tensor is the {\it symmetric} one; b) in the {\it Landau-Lifshitz} framework both the gravitational pseudo-tensor \cite{LL} and the matter tensor are the symmetric ones; c) in the {\it canonical} framework both tensors are based on the N\"other procedure and correspondingly the total energy-momentum pseudo-tensor is neither symmetric nor gauge-invariant.

In the spirit of the above references we will analyze the the dynamics of the theory at the linearized level, see {\it e.g.} \cite{BCM}, which corresponds to a perturbative treatment at first order in Newton's constant $G$. In this setting the {\it on-shell} divergent parts of the self-force of the string turn, however, out to be of order $G^2$ \cite{BD12bis}. Consequently there is an intrinsic ambiguity in the tension renormalization, inherent in standard self-force approaches, in that at first order in $G$ the divergences simply drop out. These on-shell ambiguities are absent in our approach, since we do not impose any {\it a-priori} equation of motion on the string.

With respect to the case of a string interacting minimally with a two-form potential $B_{\m\n}$, the coupling to a metric and to a dilaton introduces additional ultraviolet singularities, due to the presence in the energy-momentum tensor of  interaction-terms between the string and the fields, that are {\it localized} on the string world-sheet, see \eref{tint}. These divergences have a different origin w.r.t. the bulk-divergences of the energy-momentum tensor discussed above, and our approach entails the further advantage of separating them cleanly from the formers. This distinction is completely lost if one considers only the divergences of the total energy \cite{DH,CHH} or of the total effective action \cite{BD12} - a feature that in the past has led to conflicting results concerning tension renormalization: these contradictions are clarified and solved by our approach.

Considering gravity, as well as the exponential interactions of the dilaton, at a full non-linear level leads in the presence of distributional sources, like strings, to further problems, that we will not face, see {\it e.g.} \cite{GT}.

In the next section we present the action describing the microscopic dynamics which gives rise to self-interacting strings in $D$ space-time dimensions, and present the relevant gravitational energy-momentum pseudo-tensors. In section \ref{ld} we linearize the dynamics, restricting correspondingly the energy-momentum tensors of the fields to their quadratic expressions, and present the solutions of the linearized equations
of motion in terms of Green functions. In section \ref{cra} we introduce a {\it universal} covariant ultraviolet regularization, preserving all fundamental symmetries, and present our general approach for the derivation of the self-force.

In section \ref{siu} we apply this approach to strings in uniform motion, constructing a regularized energy-momentum tensor and performing its renormalization via subtraction of  divergent counterterms, relying on the Dirac framework. Particular attention will be paid to the cancelation of ultraviolet divergences in the {\it fundamental string} model, that comprises the non-renormalization of the string tension. This latter property, in turn, is directly related to the non-renormalization theorems of superstring amplitudes \cite{DH}, that are supposed to hold at all orders of perturbation theory. We find that, while in the Landau-Lifshitz and canonical frameworks for all $D\ge 4$ all divergences cancel, so that in particular the string tension gets not renormalized, in the Dirac framework these cancelations occur only for $D=4$. This may signal a conflict between this classical framework and the postulates of superstring theory. The subsection \ref{emc} is devoted specifically to the energy-momentum-conservation paradigm and the role of finite  counterterms in establishing the correct self-force - which for strings in uniform motion must vanish.

Sections \ref{llf} and \ref{cfw} are devoted respectively to the analogous analysis in the Landau-Lifshitz and canonical frameworks: while, as anticipated above, the actual cancelation of divergences depends on the choice of the framework, our general renormalization approach applies of course independently of the occurrence of those cancelations. In these sections we establish also  the relations between our approach and the energy-divergences analysis of  \cite{DH,CHH} and the effective-action approach of \cite{BD12}.
In section \ref{tgc} we outline the steps to be carried out in the future to derive the dynamics of self-interacting strings in arbitrary motion and discuss the uniqueness properties of our approach. This more ambitious program of using our approach to compute the self-force explicitly and compare it, where possible, with known results, may shed new light on classical-string radiation reaction, on the causality issue and, may be more marginally, on the viability of cosmic string dynamics. The final section \ref{cr} contains a concise summary of our results and of possible future developments.

\section{Classical string dynamics}\label{aad}

We consider a classical string theory in $D$ space-time dimensions whose microscopic dynamics is determined by the action
\be
\label{fands}
I=I_f+I_s,
\ee
where the field-action $I_f$ and the string-action $I_s$ are given respectively by
\begin{align}\label{ib}
I_f &=\frac{1}{G}\int \left(-R+\frac{1}{12}\,e^{-2\a\Phi}H^{\m\n\rho}H_{\m\n\rho}+\frac12\, g^{\m\n}\pa_\m\Phi\, \pa_\n\Phi\right)\sqrt{g}\,d^Dx,\\
\label{is}I_s &=-M \int e^{\bt \Phi} \sqrt{\G}\,d^2\s -\frac{\La}{2} \int W^{\m\n}B_{\m\n} \sqrt{\G}\,d^2\s.
\end{align}
We use indices $\m,\n=0,\cdots,D-1$ for the {\it bulk} space-time coordinates $x^\m$, with a mostly minus lorentzian signature, and indices $i,j=0,1$ for the {\it world-sheet} coordinates $\s^i$.  The action $I$ is inspired by superstring theory in that it corresponds to the bosonic part of the low energy effective action of ten-dimensional $N=1$ supergravity, compactified to $D$ dimensions, in the Einstein frame \cite{CFMP,GS,LTZ}. Correspondingly the space-time fields to which the string couples are the dilaton $\Phi(x)$, the axion $B_{\m\n}(x)$ and the $D$-dimensional  metric $g_{\m\n}(x)$. In \eref{ib} $R$ is the scalar curvature associated to $g_{\m\n}$ and $H_{\m\n\rho}=3\pa_{[\m}B_{\n\rho]}$ is the field strength of the axion. $G$ is related to Newton's constant through $G_N=G/16\pi$.

In the string-action \eref{is} - that describes the string propagation as well as its interaction with the bulk fields - we introduced the string coordinates $y^\m(\s)$,  with tangent vectors $U^\m_i(\s)=\pa_i y^\m(\s)$, and the induced world-sheet metric
\be\label{wsm}
\Gamma_{ij}=U^\m_i U^\n_j g_{\m\n},
\ee
with inverse $\G^{ij}$. We introduced also the antisymmetric world-sheet tensor
\be\label{w2}
W^{\m\n}=\frac{\ve^{ij}}{\sqrt{\G}}\,U^\m_i U^\n_j, \quad\mbox{where} \quad \G=-det (\G_{ij}).
\ee
On the world-sheet the space-time metric can be decomposed in parallel and orthogonal projectors
\be\label{gdec}
g^{\m\n}=L^{\m\n}+K^{\m\n}, \quad\quad L^{\m\n}=U^\m_i U^\n_j \G^{ij}.
\ee
The parallel projector $L^{\m\n}$ is sometimes referred to as the {\it first fundamental tensor}. Bulk indices and world-sheet indices are raised and lowered respectively with the metrics  $g_{\m\n}$ and $\G_{ij}$ and their inverses.

By definition, the dimensionless parameters $\a$ and $\bt$ and the dimension-one parameters $M$ and $\La$, respectively the {\it tension} and the {\it charge} of the string, are arbitrary in the {\it general} model. As we anticipated in the introduction, we will pay particular attention to the {\it fundamental string} model where they assume the values \cite{CFMP}
\be\label{fundpar}
\a=\bt=\sqrt{\frac{2}{D-2}}, \quad\quad  M=\La.
\ee
This will allow us on one hand to probe the non-renormalization properties of a superstring-inspired model \cite{DH,CHH,BC2,BD12}, and on the other to analyze the consistency  and renormalizability properties of a generic self-interacting classical string model.

Inspired by superstring theory we will assume that the dilaton takes generically a non-vanishing vacuum expectation value  $\langle\Phi\rangle \equiv \Psi $, so that, denoting its fluctuation by $\vp$, we have
\be\label{dil}
\Phi=\Psi+\vp, \quad\quad \langle\vp\rangle=0.
\ee

\subsection{Equations of motion}

The equations of motion for $\Phi$, $B_{\m\n}$, $g_{\m\n}$ and the string coordinates $y^\m$ arising from the action $I=I_f+I_s$ are
\begin{align}
g^{\m\n}D_\m \pa_\n \Phi+\frac{\a}{6}\,e^{-2\a\Phi}H^{\m\n\rho}H_{\m\n\rho}&= -GM\bt \int e^{\bt \Phi}\,\frac{\dl^D(x-y)}{\sqrt{g}}\,\sqrt{\G}\,d^2\s,\label{eqfi}\\
D_\m\left(e^{-2\a\Phi}H^{\m\n\rho}\right)&=-G\La\int W^{\n\rho}\, \frac{\dl^D(x-y)}{\sqrt{g}}\sqrt{\G}\,d^2\s,\label{eqax}\\
G^{\m\n}\equiv R^{\m\n}-\frac{1}{2}\,g^{\m\n}R&=\frac{G}{2}\, \Theta^{\m\n},\label{eqg}\\
Me^{\bt\Phi}\left(D_iU^{\m i}-\bt K^{\m\n}\pa_\n \Phi\right)&=\frac{\La}{2}\,H^{\m\n\rho}W_{\n\rho}.
\label{eqst}
\end{align}
$G^{\m\n}$ is the Einstein tensor built with the metric $g_{\m\n}$ and the generalized acceleration $D_iU^{\m i}$ of the string coordinates $y^\m$ is given by
\be\label{acc}
D_iU^{\m i}=\frac{1}{\sqrt{\G}}\,\pa_i\left(\sqrt{\G}\,\G^{ij}U_j^\m\right)+\G^\m_{\n\rho}
L^{\n\rho},
\ee
where $\G^\m_{\nu\rho}$ is the affine connection built with $g_{\m\n}$.

The {\it matter} energy-momentum tensor $\Theta^{\m\n}$ decomposes into a bulk contribution, due to the fields $\Phi$ and $B$, and a string contribution, supported on the world-sheet,
\be\label{sumsb}
\Theta^{\m\n}=\Theta^{\m\n}_b+\Theta_s^{\m\n},
\ee
given by
\begin{align}
\Theta_b^{\m\n}&=\frac{1}{G}\left(D^\m\Phi D^\n\Phi-\frac{1}{2}\,g^{\m\n}D^\rho\Phi D_\rho\Phi +\frac{1}{2}\,e^{-2\a\Phi}\left(H^{\m\rho\s}H^\n{}_{\rho\s}-\frac{1}{6}\,g^{\m\n}
H^{\rho\s\la}H_{\rho\s\la}\right)\right)\label{enfb},\\
\Theta^{\m\n}_s&=M\int \label{ts} e^{\bt\Phi}L^{\m\n}\,\frac{\dl^D(x-y)}{\sqrt{g}}\sqrt{\G}\,d^2\s.
\end{align}
Obviously in $\Theta^{\m\n}_s $ there is no contribution from the axion field $B_{\m\n}$ because its minimal coupling to the string in \eref{is}, being topological, does not contain the metric.

Computing {\it mechanically} the  covariant divergence of $\Theta^{\m\n}$ one obtains the identity
\be
\begin{aligned}\label{consform}
D_\m\Theta^{\m\n}=&\frac{1}{2G}H^\n{}_{\rho\s}\,D_\m\left(e^{-2\a\Phi}H^{\m\rho\s}\right)
+\frac{1}{G}\left(g^{\rho\s}D_\rho\pa_\s\Phi+\frac{\a}{6}\,
e^{-2\a\Phi}H^{\rho\s\la}H_{\rho\s\la}\right)D^\n\Phi\\
&+M\int e^{\bt\Phi}\left(D_iU^{\n i}+\bt L^{\n\m}\pa_\m\Phi\right)
\frac{\dl^D(x-y)}{\sqrt{g}}\,\sqrt{\G}\,d^2\s,
\end{aligned}
\ee
and if one uses the equations \eref{eqfi}, \eref{eqax} and \eref{eqst} one gets obviously $D_\m\Theta^{\m\n}=0$. As stressed in the introduction, the operations leading to \eref{consform} have however only {\it formal} validity, in that $\Theta^{\m\n}$ is not a distribution - it diverges too violently in the vicinity of the string - and hence its $D$-divergence  ``$\pa_\m \Theta^{\m\n}$'' is meaningless. {\it A fortiori} one is not allowed to resort to the Leibnitz-rule $\pa_\mu(f_1f_2)=\pa_\m f _1 f_2+f_1\pa_\m f_2$, that has been used thoughtless to derive \eref{consform}.

\subsection{Gravitational energy-momentum pseudo-tensors}

Since the implementation of $D$-momentum conservation requires an energy-momentum tensor that satisfies a {\it standard} continuity equation, before attacking the renormalization issue we must recast the formal equation $D_\m\Theta^{\m\n}=0$ in a (still formal) equation of the type $\pa_\m \tau^{\m\n}=0$, for some pseudo-tensor $\tau^{\m\n}$. A standard continuity equation is also in line with our distributional framework in that the $D$-divergence of a distribution - as $\tau^{\m\n}$ should eventually be - is {\it always} a distribution, while on the contrary an object like $D_\m \Theta^{\m\n} \sim \pa \Theta +\G \Theta$ - involving {\it products} between distributions - would not be so.

To attack this problem we must face first the issue of the - non unique - {\it gravitational} energy-momentum pseudo-tensor. We resort to three different choices, giving rise to the three different conservation {\it frameworks} described in the introduction.

\subsubsection{Dirac's energy-momentum pseudo-tensor}

The distinguished feature of Dirac's gravitational energy-momentum pseudo-tensor $\Sigma^{\m}{}_\n$  \cite{PAM} is that it descends canonically from N\"other's theorem, applied to the Einstein-Hilbert action. It carries one upper and one lower index and reads
\be\label{TD}
\Sigma^{\m}{}_\n=
\frac{1}{G}\left(\left(\G^\m_{\a\bt}-\delta^\m_\a\G_{\bt\la}^\la\right)\pa_\nu\left(\sqrt{g}
g^{\a\bt}\right) -\frac{1}{2}\,\dl^\m{}_\n
\left(\G^\g_{\a\bt}-\delta^\g_\a\G_{\bt\la}^\la\right)\pa_\g\left(\sqrt{g}
g^{\a\bt}\right)\right).
\ee
Notice that $\Sigma^{\m}{}_\n$ is quadratic in the first derivatives of the metric. The term multiplying $\dl^\m{}_\n$  is related to the Einstein-Hilbert action through the identity
\[
-\frac{1}{G} \int\! R\sqrt{g}\,d^Dx=\frac{1}{2G}\int\! \left(\G^\g_{\a\bt}-\delta^\g_\a\G_{\bt\la}^\la\right)\pa_\g\left(\sqrt{g}
g^{\a\bt}\right) d^Dx,
\]
{\it i.e.} it differs from $R\sqrt{g}$ by total derivatives, and represents thus an equivalent quadratic lagrangian.

As shown by Dirac, $\Sigma^{\m}{}_\n$ is tied to the Einstein tensor $G^{\m\n}=R^{\m\n}-\frac{1}{2}\,g^{\m\n}R$ through the identity
\be\label{idenD}
\pa_\m\left(\Sigma^{\m}{}_\nu+\frac{2}{G}\,\sqrt{g}\,G^\m{}_\nu\right)=0.
\ee
Introducing the {\it total} energy-momentum tensor - actually a pseudo-tensor, too - with one upper and one lower index
\be\label{tott}
\tau^\m{}_\n\equiv\sqrt{g}\,\Theta^\m{}_\n+\Sigma^{\m}{}_\nu,
\ee
from \eref{idenD} and $D_\m G^\m{}_\n=0$ we derive that it satisfies the identity
\be\label{iden0}
\pa_\m \tau^\m{}_\n= \sqrt{g}\,D_\m \Theta^{\m}{}_\nu-\frac{\sqrt{g}}{G} \left(G^{\a\bt}-\frac{G}{2} \,\Theta^{\a\bt}\right)\pa_\n g_{\a\bt}.
\ee
Since the matter energy-momentum tensor $\Theta^\m{}_\n$ satisfies the identity \eref{consform}, $\tau^\m{}_\n$ obeys the continuity equation $\pa_\m \tau^\m{}_\n=0$, if all fields satisfy their equations of motion \eref{eqfi}-\eref{eqst}.

Since \eref{idenD} is an algebraic identity we infer  the existence of a tensor $P^{\rho\m}{}_\nu$ - antisymmetric in $\rho$ and $\m$ and built only with $g_{\m\n}$ - such that
\be\label{d1}
\sqrt{g}\,G^\m{}_\nu=\frac{G}{2}\,\big(\pa_\rho P^{\rho\m}{}_\nu-\Sigma^{\m}{}_\nu\big).
\ee
A direct calculation gives\footnote{The most efficient way to perform it is to extract from $\sqrt{g}\,G^\m{}_\nu$ all terms linear in $\pa\pa g$ and to cast them in the form of a divergence of an antisymmetric tensor. As in the whole paper in \eref{wex} antisymmetrization is understood with unit weight.}
\be\label{wex}
P^{\rho\m}{}_\nu=\frac{2}{G\sqrt{g}}\,
\pa_\bt\left(gg^{\g[\m}g^{\rho]\bt}\right)g_{\g\n}.
\ee
Actually equations \eref{d1} and \eref{wex} could be taken equivalently as the defining equations for $\Sigma^\m{}_\n$.

\subsubsection{The Landau-Lifshitz energy-momentum pseudo-tensor}

In analogy to \eref{d1} and \eref{wex}, the Landau-Lifshitz gravitational energy-momentum pseudo-tensor $\w\Sigma^{\m\n}$   \cite{LL} - a {\it symmetric} tensor with two upper indices - is defined through the relations
\be\label{ll}
gG^{\m\n}=\frac{G}{2}\,\big(\pa_\rho \w P^{\rho\m\n}-\w\Sigma^{\m\n}\big),
\ee
where the tensor $\w P^{\rho\m\n}$,  antisymmetric in $\rho$ and $\m$, is by definition \cite{LL}
\be\label{z}
\w P^{\rho\m\n}=
\sqrt g\, P^{\rho\m}{}_\g g^{\g\n}=
\frac{2}{G}\,
\pa_\bt\left(gg^{\n[\m}g^{\rho]\bt}\right).
\ee
Like $\Sigma^\m{}_\n$ also $\w\Sigma^{\m\n}$ can be seen to be quadratic in the first derivatives of the metric, and from \eref{ll} follows the identity
\[
\pa_\m\left(\w\Sigma^{\m\n}+\frac{2}{G}\,gG^{\m\n}\right)=0,
\]
analogous to \eref{idenD}. In this framework the {\it total} energy-momentum tensor, with two upper indices, is defined by
\be\label{ttilde}
\w\tau^{\m\n}\equiv g\Theta^{\m\n}+\w\Sigma^{\m\n},
\ee
and thanks to $D_\m G^{\m\n}=0$ this time one arrives at
\be\label{iden1}
\pa_\m \w\tau^{\m\n}= g D_\m \Theta^{\m\n}+\frac{2\sqrt{g}}{G}\left(G^{\a\bt}-\frac{G}{2} \,\Theta^{\a\bt}\right)\left(\sqrt{g}\,\G^\n_{\a\bt}-\dl_\a^\n\pa_\bt\sqrt{g}\right), \ee
counterpart of \eref{iden0}. The r.h.s.  vanishes again if the equations \eref{eqfi}-\eref{eqst} hold.

With the help of \eref{z} we can establish an explicit link between $\tau^\m{}_\n$ and $\w \tau^{\m\n}$. Equating the r.h.s. of \eref{ll} with $\sqrt{g}$ times the r.h.s. of \eref{d1} with the index $\n$ raised,  we establish first the  link between the pseudo-tensors $\w\Sigma^{\m\n}$ and $\Sigma^\m{}_\n$
\be\label{dll}
\w\Sigma^{\m\n}=\sqrt{g}\,\Sigma^\m{}_\rho\, g^{\rho\n}+
P^{\rho\m}{}_\g\,\pa_\rho\left(\sqrt{g}g^{\g\n}\right),
\ee
that is consistent with the fact that both tensors are quadratic in $ \pa g$.
\eref{dll} implies then the relation between the total energy-momentum tensors \eref{tott} and \eref{ttilde}
\be\label{tot12}
\w \tau^{\m\n}= \sqrt{g}\,\tau^\m{}_\rho\, g^{\rho\n}+
P^{\rho\m}{}_\g\,\pa_\rho\left(\sqrt{g}g^{\g\n}\right).
\ee
From this relation, using again \eref{tott} and \eref{d1}, one finds eventually that the two total energy-momentum tensors are connected through a three-tensor $L^{\rho\m\n}$,  antisymmetric in its first two indices, modulo equations of motion, as it should be:
\be\label{t12}
\w\tau^{\m\n}=\tau^\m{}_\rho\eta^{\rho\n}+\pa_\rho L^{\rho\m\n}-\frac{2\sqrt g}{G}\left(
G^\m{}_\rho-\frac{G}{2}\,\Theta^\m{}_\rho\right)\big(\sqrt{g}g^{\rho\n}-\eta^{\rho\n}\big),
\ee
where
\be\label{l3}
L^{\rho\m\n}=P^{\rho\m}{}_\a \big(\sqrt{g}g^{\a\n}-\eta^{\a\n}\big).
\ee
The analysis of this paper will be performed primarily in the Dirac framework,  based on the energy-momentum tensor $\tau^\m{}_\n$. Equations  \eref{tot12} and \eref{t12} will then be used to translate this analysis to the Landau-Lifshitz framework, based on $\w\tau^{\m\n}$

\subsubsection{The canonical energy-momentum pseudo-tensor}

By definition the {\it canonical} total energy-momentum tensor $\wh\tau^\m{}_\n$  follows from N\"other's theorem applied to the whole action \eref{fands}. Consequently it differs from Dirac's choice \eref{tott} by the divergence of a three-tensor, antisymmetric in its first two indices, modulo the equations of motion of the axion\footnote{The dilaton is a scalar and so its canonical and symmetric energy-momentum tensors coincide.}:
\be\label{cano}
\wh\tau^\m{}_\n\equiv \tau^\m{}_\n+\pa_\rho S^{\rho\m}{}_\nu
+\frac{\sqrt g}{G}\left(\!D_\rho\left(e^{-2\a\Phi}H^{\rho\m\s}\right)+G\La\int\! W^{\m\s}\, \frac{\dl^D(x-y)}{\sqrt{g}}\sqrt{\G}\,d^2\s\right)\!B_{\s\n}.
\ee
The tensor $S^{\rho\m}{}_\n$ is quadratic in the axion and reads
\be\label{sss}
S^{\rho\m}{}_\nu=\frac{\sqrt{g}}{G}\,e^{-2\a\Phi}H^{\rho\m\s}B_{\s\n}= -S^{\m\rho}{}_\nu.
\ee
The  major shortcoming of the tensor \eref{cano} is that it is no longer gauge-invariant under $\dl B_{\m\n} =\pa_\m\La_\n -\pa_\n\La_\m$. Inserting \eref{sss} in \eref{cano} we obtain the relation
\be
\label{cano1}
\wh\tau^\m{}_\n= \tau^\m{}_\n -
\frac{\sqrt{g}}{G}\,e^{-2\a\Phi}H^{\m\rho\s}\pa_\rho B_{\s\n}-\La
\int W^{\m\s}B_{\s\n}\,\dl^D(x-y)\sqrt{\G}\,d^2\s.
\ee
The second term at its r.h.s. amounts in \eref{enfb} to the replacement
\[
H^{\m\rho\s}H_{\n\rho\s}\quad\rightarrow\quad
H^{\m\rho\s}\pa_\n B_{\rho\s},
\]
while the third term represents a modification of the world-sheet term \eref{ts}, corresponding to the replacement
\[
Me^{\bt\Phi}L^\m{}_\n\quad\ra\quad
Me^{\bt\Phi}L^\m{}_\n-\La W^{\m\s}B_{\s\n}.
\]
From \eref{iden0} and \eref{cano} we deduce that $\pa_\m\wh\tau^\m{}_\n=0$, if the fields satisfy \eref{eqfi}-\eref{eqst}.

\section{Linearized dynamics}\label{ld}

As stated previously we consider the theory in the linear regime, which corresponds to an analysis at first order in Newton's constant $G$. This amounts to keep in the field-action
\eref{ib} the terms quadratic in the fields, and in the string-action \eref{is} the terms linear in the fields. Correspondingly in the equations of motion \eref{eqfi}-\eref{eqst} we must keep only the  terms linear in the fields.

We write the dilaton as in \eref{dil} as its constant vacuum expectation value $\Psi$ plus a fluctuation $\vp$, {\it i.e.} $\Phi(x)=\Psi +\vp(x)$. Moreover,  to simplify the formalism, we parameterize the fluctuation of the space-time metric in terms of a symmetric field $F^{\m\n}$, specified by
\be\label{gf}
 g_{\m\n}=\eta_{\m\n}+h_{\m\n}\equiv  \eta_{\m\n}+F_{\m\n}-\frac{1}{D-2}\,\eta_{\m\n}F,\quad\quad F\equiv F_{\a\bt}\eta^{\a\bt},
\ee
{\it i.e.} $F_{\m\n}=h_{\m\n}-\frac{1}{2}\,\eta_{\m\n}h^\a{}_\a$ and $F=\left(1-\frac{D}{2}\right)h^\a{}_\a$.
This choice is convenient in that we have, in any dimension $D\ge4$,
\be\label{ghat}
\sqrt{g}g^{\m\n}=\eta^{\m\n}-F^{\m\n}+o(F^2).
\ee
In particular, the harmonic gauge for diffeomorphisms $\pa_\m(\sqrt{g}g^{\m\n})=0$, that we will use throughout the rest of the paper, assumes then the simple linearized form
\be\label{hamonic}
\pa_\m F^{\m\n}=0.
\ee
For the axion we use the Lorenz-gauge $\pa_\m B^{\m\n}=0$. Henceforth all indices will be raised and lowered with the flat metric $\eta_{\m\n}$.
In conclusion, the linearization will be in the fields $f=\{\vp, B_{\m\n}, F_{\m\n}\}$.

\subsection{Equations of motion}\label{eom}

The linearized equations of motion \eref{eqfi}-\eref{eqst} read $(\square =\pa_\m\pa^\m$)
\begin{align}
 \square\vp=& - e^{\bt\Psi}GM\bt \int\dl^D(x-y)\sqrt{\g}\,d^2\s,\label{eqfil}\\
 \square B_{\m\n} &=-e^{2\a\Psi} G\La\int w_{\m\n}\, \dl^D(x-y)\sqrt{\g}\,d^2\s,\label{bmn}\\
\square F_{\m\n}=&-e^{\bt\Psi} GM\int l_{\m\n}\, \dl^D(x-y)\sqrt{\g}\,d^2\s,\label{eqgl}\\
Me^{\bt\Psi} \Delta_i U^{\m i}=&\frac{\La}{2}\,H^{\m\a\bt}w_{\a\bt}+
Me^{\bt\Psi}\left[k^{\m\n}\left(\left(\frac{1}{2}\,\pa_\n F_{\a\bt}-\pa_\a F_{\bt\n}\right)l^{\a\bt}+\frac{1}{D-2}\,\pa_\n F-\bt\pa_\n\vp\right)\right.\nn\\
 &\left.l^\m{}_{\a\bt}F^{\a\bt} -k^{\m\n}F_{\n\rho}\,\Delta_i U^{\rho i}+\left(\frac{F}{D-2}-\frac{1}{2}\,l^{\a\bt}F_{\a\bt}\right)\Delta_i U^{\m i}
 \right] \equiv {\cal S}^\m_{bare},
\label{ill}
 \end{align}
where we introduced the flat-space counterparts of the tensors $\G_{ij}$, $L^{\m\n}$, $K^{\m\n}$ and $W^{\m\n}$ in \eref{wsm}-\eref{gdec}
\be\label{flat}
\g_{ij}=U_i^\m U_j^\n\eta_{\m\n}, \quad \g=-det \g_{ij},\quad l^{\m\n}=\,U^\m_i U^\n_j\g^{ij},\quad k^{\m\n}=\eta^{\m\n}-l^{\m\n}, \quad w^{\m\n}=\frac{\ve^{ij}}{\sqrt{\g}}\,U^\m_i U^\n_j,
\ee
obeying the relations
\be\label{proj}
l^{\m\rho}l_{\rho\n}=l^\m{}_\n,\quad\quad k^{\m\rho}k_{\rho\n}=k^\m{}_\n,\quad\quad
k^{\m\rho}l_{\rho\n}=0,\quad\quad w^{\a\bt}w^{\m\n}=l^{\a\n}l^{\bt\m}-l^{\a\m}l^{\bt\n}.
\ee
Correspondingly $l^{\m}{}_{\m}=2$ and $k^\m{}_\m=D-2$. $\Delta_i$ is the covariant derivative w.r.t. the flat-target-space world-sheet metric $\g_{ij}$ and $\Delta_i U^{\m i}$ is the flat-target-space reparameterization invariant acceleration of the string
\be\label{acclin}
\Delta_i U^{\m i}=\frac{1}{\sqrt{\g}}\,\pa_i\left(\sqrt{\g}\,\g^{ij}U_j^\m\right).
\ee
In \eref{ill} $l^\m{}_{\a\bt}$ represents the {\it second fundamental tensor}, which can be expressed in different ways and entails several properties {\it e.g.} ($\Delta_\a\equiv U_\a^i\Delta_i$)
\be\label{2fund}
l^\m{}_{\a\bt}=l_{\g\bt}\,\Delta_\a l^{\g\m}=k^{\m\g}\,\Delta_\a l_{\g\bt},\quad\quad l^\m{}_{\a\bt}=l^\m{}_{\bt\a}, \quad\quad   l_{\n \m}\,l^\m{}_{\a\bt}=0.
\ee

Equations \eref{eqgl} are the linearized Einstein equations. Applying $\pa_\m$ to both sides of them one arrives formally at a mismatch, since at its r.h.s. one gets a non-vanishing term: $\square \pa_\m F^{\m\n}\sim G\Delta_iU^{\n i}\neq 0$. This is a remnant of the peculiar property of Einstein's equations to {\it imply} the geodesic equation of motion, in the present case the string equation of motion. At the linearized level this does however not lead to an inconsistency; in fact, since $\Delta_iU^{\n i}$ eventually equals the finite self-force - which is of order $G$ - the above mismatch is of order  $o(G^2)$.

Just for the sake of completeness above we wrote out also the linearized version \eref{ill} of the string equation of motion \eref{eqst} that, contrary to the field equations \eref{eqfil}-\eref{eqgl}, is actually ill-defined. In fact, in \eref{ill} the fields $f(x)$
are evaluated at the world-sheet $x=y(\sigma)$, where they diverge, so that the bare self-force ${\cal S}^\m_{bare}$ is infinite. There is moreover an intrinsic ambiguity in this equation concerning the renormalization of the string tension $M$, due to the appearance of the acceleration $\Delta_i U^{\m i}$ also at its right hand side, where it is multiplied by the self-fields $f(y(\s))$. Since the latter are of order $G$, as is the acceleration $\Delta_i U^{\m i}$, the terms of the kind $\Delta_i U^{\m i}f(y(\s))$ are actually of order $G^2$ and should have therefore be omitted in \eref{ill} from the beginning. Similarly (the divergent parts of) the terms $\pa f(y(\s)) $ in the first line of \eref{ill} have the structure $\pa f(y(\s))\sim G\Delta_i U^{\m i}\sim G^2$ \cite{BC2,BCM}. Consequently, as observed already in \cite{BD12bis}, in a first-order setting as the present one, it appears intrinsically impossible to perform an unambiguous {\it quantitative} analysis of tension renormalization, {\it upon renormalizing directly ${\cal S}^\m_{bare}$ at the basis of} \eref{ill}.

According to our approach, our starting point to derive a finite self-force will actually not be \eref{ill}, but rather energy-momentum conservation.

\subsection{Linearized total energy-momentum tensors}

\subsubsection{Dirac framework}

We present first the linearized version of the total energy-momentum tensor \eref{tott} of the  Dirac framework. We write it as a sum of three terms, each term having its specific physical meaning. For notational convenience we write it with two flat {\it upper} indices - an operation that preserves the formal conservation law $\pa_\m \tau^{\m\n}=0$ - albeit maintaining for the linearized version the same symbol $\tau$ as for the exact one:
\be\label{123}
\tau^{\m\n}\equiv \tau^\m{}_\rho\big|_{lin}\,\eta^{\rho\n} =\tau_f^{\m\n} +\tau_{int}^{\m\n}+\tau_{\rm kin}^{\m\n}.
\ee
The first term, the {\it field} energy-momentum tensor $\tau_f^{\m\n}$, represents the energy due solely to the fields and is supported on the bulk. It is obtained extracting from \eref{enfb} and \eref{TD}
the terms quadratic in the fields $f=\{\vp,B,F\}$
\be
\begin{aligned}
\label{tf}
\tau^{\m\n}_f=& \left(\sqrt{g}\,\Theta_b^{\m\bt}{}g_{\bt\rho}+\Sigma^\m{}_\rho\right)\Big|_{f^2}\,
\eta^{\rho\n}\\
=&\frac{1}{G}\left\{\pa^\m\vp\,\pa^\n\vp-\frac{1}{2}\,\eta^{\m\n}\pa^\rho
\vp\pa_\rho\vp+\frac{1}{2}\,e^{-2\a\Psi}\left(H^{\m\a\bt}H^\n{}_{\a\bt}
-\frac{1}{6}\,\eta^{\m\n}
H^{\a\bt\g}H_{\a\bt\g}\right)\right.\\
+&\frac{1}{2}\,\pa^\m F^{\a\bt}\pa^\n F_{\a\bt}-\pa^\a F^{\bt\m}\pa^\n F_{\a\bt}  -\frac{1}{2(D-2)}\,\pa^\m F\pa^\n F\\
-&\left.\frac{1}{2}\,\eta^{\m\n}\left(\frac{1}{2}\,\pa^\g F^{\a\bt}\pa_\g F_{\a\bt}-\pa^\a F^{\bt\g}\pa_\g F_{\a\bt}  -\frac{1}{2(D-2)}\,\pa^\g F\pa_\g F\right)\right\}.
\end{aligned}
\ee
The gravitational contribution of $\tau^{\m\n}_f$ in the last two rows is not symmetric in its indices - a characteristic feature of the Dirac tensor \eref{TD}.

The {\it interaction} energy-momentum tensor $\tau^{\m\n}_{int}$ arises from the interaction between the fields and the string and is hence supported on the world-sheet. It is obtained extracting from $\sqrt{g}\,\Theta^{\m\bt}_sg_{\bt\rho}$, see \eref{ts}, the terms linear in the fields:
\be
\begin{aligned}
\tau^{\m\n}_{int}&=\left(\sqrt{g}\,\Theta^{\m\bt}_sg_{\bt\rho}\right)
\big|_f\, \eta^{\rho\n}\\
&=Me^{\bt\Psi}\int\left(\left(\frac{1}{2}\,l^{\a\bt}F_{\a\bt}-\frac{F}{D-2}
+\bt\vp\right)l^{\m\n}-l^{\m\a}l^{\n\bt}F_{\a\bt}+l^{\m\a} F_\a{}^\n\right)\dl^D(x-y)\sqrt{\g}\,d^2\s.\label{tint}
\end{aligned}
\ee
Only the gravitational field and the dilaton contribute to $\tau^{\m\n}_{int}$, but not the axion, for the reasons explained above.

The term $\tau_{\rm kin}^{\m\n}$ represents the free {\it kinetic} energy-momentum tensor of the string and is obtained from $\Theta_s^{\m\n}$ \eref{ts} setting all fields $f=\{\vp,B,F\}$ to their background values, {\it i.e.} zero,
\be\label{tslin}
\tau_{\rm kin}^{\m\n}=\left(\sqrt{g}\,\Theta_s^{\m\bt}g_{\bt\rho}\right)
\big|_{f=0}\,\eta^{\rho\n}=Me^{\bt\Psi}\int l^{\m\n}\dl^D(x-y)\sqrt{\g}\,d^2\s.
\ee

\subsubsection{Landau-Lifshitz framework}

In the framework of Landau and Lifshitz the linearized equations of motion \eref{eqfil}-\eref{eqgl} remain clearly the same, what changes is the form of the
energy-momentum tensor $\w \tau^{\m\n}$ in \eref{ttilde}. The most simplest way to write it down is to use its relation to the Dirac-tensor \eref{tot12}. Setting as in \eref{123} - from now on with the symbol $\w\tau^{\m\n}$ we understand  its linearized version -
\be\label{123ll}
\w \tau^{\m\n}=\w\tau_f^{\m\n}+\w\tau_{int}^{\m\n}+\tau_{\rm kin}^{\m\n},
\ee
from \eref{tot12} we see that $\w\tau_f^{\m\n}$ receives  additional contributions from the gravitational field and that, due to the presence of the factor $\sqrt{g}g^{\rho\n}$ in the first term of \eref{tot12}, also the interaction-term changes:
\begin{align}\label{fdll}
\w\tau^{\m\n}_f&=\tau_f^{\m\n}+ \frac{1}{G}\left(\pa_\a F_\bt{}^\m \pa^\a
F^{\bt\nu}+\pa_\a F^{\a\bt}\pa_\bt F^{\m\n}-\pa_\a F^{\a\m}\pa_\bt F^{\bt\n}-\pa^\a F^{\bt\n}\pa^\m F_{\a\bt} \right),\\
\w\tau^{\m\n}_{int}&=\tau^{\m\n}_{int}-
Me^{\bt\Psi}\int l^{\m\a} F_\a{}^\n\,\dl^D(x-y)\sqrt{\g}\,d^2\s.\label{intll}
\end{align}
To derive \eref{fdll} we used in particular the linearized version of \eref{wex}
\be\label{wexl}
P^{\rho\m}{}_\n=\frac{2}{G}\left(\pa^{[\m}F^{\rho]}{}_\n+\pa_\bt F^{\bt[\m}
\dl_\n^{\rho]}\right),
\ee
following from \eref{ghat}. Obviously $\tau_{\rm kin}^{\m\n}$, the free energy-momentum tensor of the string, remains the same. It is easily seen that the tensor $\w \tau^{\m\n}$ given by \eref{123ll}-\eref{intll} is symmetric.

\subsubsection{Canonical framework}\label{cf}

Writing also the linearized version of the canonical total energy-momentum tensor in the form
\be\label{canlin}
\wh \tau^{\m\n}=\wh\tau_f^{\m\n}+\wh\tau_{int}^{\m\n}+\tau_{\rm kin}^{\m\n},
\ee
from \eref{cano1} we obtain
\begin{align}\label{fdcan}
\wh\tau^{\m\n}_f&=\tau_f^{\m\n}-\frac{1}{G}\,e^{-2\a\Psi}H^{\m\rho\s}\pa_\rho B_\s{}^\n,
\\
\wh\tau^{\m\n}_{int}&=\tau^{\m\n}_{int}-\La
\int w^{\m\s}B_\s{}^\n\,\dl^D(x-y)\sqrt{\g}\,d^2\s.\label{intcan}
\end{align}
Thanks to (the linearized versions of) \eref{iden0}, \eref{iden1} and \eref{cano}, the {\it formal} conservation laws
\be\label{conslin}
\pa_\m\tau^{\m\n}=0, \quad\quad\pa_\m\w\tau^{\m\n}=0, \quad\quad\pa_\m\wh\tau^{\m\n}=0
\ee
hold, if the fields and the string coordinates satisfy the equations of motion \eref{eqfil}-\eref{ill}. Since singularities do arise only on the world-sheet, and the tensors
$\tau_{int}^{\m\n}$, $\w\tau_{int}^{\m\n}$, $\wh\tau_{int}^{\m\n}$, as well as $\tau_{\rm kin}^{\m\n}$,  are supported on the world-sheet, too, the formal equations \eref{conslin} imply that, if only the fields satisfy their equations of motion \eref{eqfil}-\eref{eqgl}, in the {\it complement of the world-sheet} the field tensors satisfies the {\it true} conservation laws
\be\label{fieldt}
\pa_\m\tau^{\m\n}_f=0, \quad\quad\pa_\m\w\tau^{\m\n}_f=0, \quad\quad\pa_\m\wh\tau^{\m\n}_f=0.
\ee
This property will become crucial later one.

\section{Regularized field solutions and renormalization}\label{cra}

We address now the solutions of the equations \eref{eqfil}-\eref{eqgl} obeyed by the fields  $f=\{\vp,B,F\}$.
They are all of the d'Alembert-type
\be\label{square}
\square f(x)=\int j(\s)\,\dl^D(x-y(\s))\,d^2\s
\ee
and admit thus the solution
\be\label{fx}
f(x)=\int {\cal G}(x-y(\s)) j(\s)\,d^2\s,
\ee
where ${\cal G}(x)$ - the retarded Green function in $D$ space-time dimensions - satisfies the equation $\square {\cal G}(x)=\dl^D(x)$; for explicit expressions see \eref{greene} with $\ve=0$.

\subsection{Singularities and distributions}

Denoting generically the $D-2$ coordinates orthogonal to the string world-sheet with $x_\perp$, in the vicinity of the string, that is for $x_{\perp}\ra 0$, the fields \eref{fx} diverge schematically as (see for example \eref{fe1}-\eref{fe3} with $\ve=0$)
\be
\label{fxdiv}
f(x)\sim\begin{cases}
\dis \frac{1}{x_\perp^{D-4}},&\mbox{for }D>4,\\
 \ln x_\perp,&\mbox{for }D=4.
 \end{cases}
\ee
The behaviors \eref{fxdiv} represent {\it distributional} types of singularity: the fields $f(x)$ in \eref{fx} {\it are} indeed distributions.
In contrast the {\it bare} field energy-momentum tensor \eref{tf} diverges for $x_{\perp}\ra 0$ as
\be\label{tmndiv}
\tau^{\m\n}_f(x)\sim \pa f(x)\pa f(x)\sim \frac{1}{x_\perp^{2D-6}},
\ee
a behavior that is {\it not} of the distributional type\footnote{Applying \eref{tmndiv} to a test function $\vp(x)= \vp(x_\perp,x^0,x^1)$, schematically one has
\[
\int \tau^{\m\n}_f(x)\,\vp(x)\,d^{D-2}x_\perp dx^0dx^1 \sim \int\frac{dx_\perp}{x_\perp^{D-3}},
\]
that diverges for $D\ge4$.}, unless $D<4$. Said in other words, for $D\ge 4$ the functions $\tau^{\m\n}_f(x)$ are not distributions.

Similarly, also the interaction energy-momentum tensor $\tau_{int}^{\m\n}$ \eref{tint} is ill-defined, because the self-fields $f(y(\s))=f(x)|_{x_\perp =0}$ appearing therein are infinite. Contrary to the singularities of $\tau^{\m\n}_f$, the singularities of $\tau_{int}^{\m\n}$  are hence {\it strongly} local, {\it i.e.} they are localized on the world-sheet as is the whole  $\tau_{int}^{\m\n}$. Correspondingly their subtraction encounters no technical difficulty, so that the finite part of $\tau_{int}^{\m\n}$ gives rise directly to a {\it renormalized} interaction energy-momentum tensor $T^{\m\n}_{int}$, see \eref{tmnint} below.

By contrast the construction of a renormalized field energy-momentum tensor $T^{\m\n}_f$ out of $\tau_f^{\m\n}$ is more involved. We impose on $T^{\m\n}_f$ the {\it minimal} conditions:
\begin{itemize}
\item[a)] $T^{\m\n}_f(x)$ is a tempered distribution, {\it i.e.} an element of ${\cal S}'(\mathbb{R}^D)$;
\item[b)] $T^{\m\n}_f(x)=\tau^{\m\n}_f(x)$, if $x$ belongs to the complement of the world-sheet.
\end{itemize}
Condition a) is a necessary {\it pre}-consistency condition for local energy-momentum conservation: the distributional divergence  $\pa_\m T^{\m\n}_f$ of a distribution is indeed always a distribution. Condition b) says instead that we want to modify $\tau_f^{\m\n}$ ``as little as possible'', {\it i.e.} we do not want to change its values in the complement of the world-sheet, since there it is regular. This condition represents a cornerstone of our approach.

By construction conditions a) and b) determine  $T^{\m\n}_f$ modulo terms supported on the world-sheet: this is the aforementioned {\it finite-counterterm-ambiguity}, that we have to take into account in the following.

\subsection{Covariant regularization and renormalized energy-momentum tensor}\label{rem}

To construct out of the (ill-defined) tensor $\tau_f^{\m\n}$ a tensor $T^{\m\n}_f$ satisfying the above conditions a) and b), we need first of all a set of {\it regular} fields $f_\ve(x)$, that for $\ve\ra0$ tend (pointwise and in the sense of distributions) to the fields $\eref{fx}$. In what follows $\ve$ is a positive regularization parameter with the dimension of a length. A convenient covariant regularization consists in replacing in the solutions \eref{fx} the Green functions ${\cal G}(x)$ with the regularized - but still Lorentz-invariant - Green functions (for $D=4$ see \cite{LR})
\be
{\cal G}_\ve(x)=
\begin{cases}{\displaystyle
{H(x^0)\over 2\pi^{N+1}} \,\left({d\over dx^2}\right)^N\dl(x^2-\ve^2)},&\mbox{for }D=2N+4, \\
&\label{greene}\\
{\displaystyle{H(x^0)\over 2\pi^{N+1}} \,\left({d\over dx^2}\right)^N{H(x^2-\ve^2)\over \sqrt{x^2-\ve^2}}},&\mbox{for }D=2N+3,
\end{cases}
\ee
where $H(\,\cdot\,)$ is the Heaviside function and $x^2=x_\m x^\m$. In practical applications of these formulae it may be useful to replace the derivative $d/dx^2$ with $-d/d\ve^2$. The smoothed fields
\be
\label{fxe}
f_\ve(x)=\int {\cal G}_\ve(x-y(\s)) j(\s)\,d^2\s
\ee
are now {\it regular} distributions and on the world-sheet one has in particular the small-$\ve$ behaviors, see below,
\be
\begin{aligned}
f_\ve(y(\s))\sim\begin{cases}\displaystyle\frac{1}{\ve^{D-4}}, & \mbox{for }D>4,\\
\ln\ve, & \mbox{for }D=4.
\end{cases}
\end{aligned}
\ee

The main virtue of the regularization \eref{greene} is that it preserves manifest Lorentz- as well as reparameterization-invariance. Consequently the regularized field energy-momentum tensor $\tau_{f\ve}^{\m\n}$ - obtained from \eref{tf} replacing the fields $f$ with $f_\ve$ - are distributions, too, and they are covariant {\it tensors}. However, while in the complement of the world-sheet one has the  {\it point-wise} limit
\[
\lim_{\ve\ra 0}\tau_{f\ve}^{\m\n}(x)=\tau_{f}^{\m\n}(x),
\]
the {\it distributional} limit
\[
 {\cal S}'-\lim_{\ve\ra0}\tau_{f\ve}^{\m\n}
\]
does not exist. Indeed, before taking this limit one must isolate from $\tau_{f\ve}^{\m\n}$ the term $\tau_{f\ve}^{\m\n}\big|_{div}$ that diverges as $\ve\ra0$ and that, in turn, must be supported on the world-sheet. The renormalized energy-momentum tensor $T^{\m\n}_f$ can then be defined subtracting this {\it divergent counterterm} and performing then the distributional limit
\be\label{tmnf}
T^{\m\n}_f\equiv{\cal S}'-\lim_{\ve\ra0}\left(\tau_{f\ve}^{\m\n}-\tau_{f\ve}^{\m\n}\big|_{div}\right).
\ee
By construction this tensor satisfies the above conditions a) and b).

Similarly one introduces a regularized interaction energy-momentum tensor $\tau_{int\,\ve}^{\m\n}$, replacing in \eref{tint} the fields $f$ with $f_\ve$, and subtracts then its divergent part obtaining the renormalized  interaction energy-momentum tensor
\be\label{tmnint}
T^{\m\n}_{int}\equiv{\cal S}'-\lim_{\ve\ra0}\left(\tau_{int\,\ve}^{\m\n}-\tau_{int\,\ve}^{\m\n}\big|_{div}\right).
\ee
Although the formulae \eref{tmnf} and \eref{tmnint} are formally identical,  in \eref{tmnint} the subtraction of the divergent counterterm, as anticipated above, will be a conceptually {\it trivial} operation, while the analogous process in \eref{tmnf} will require the whole apparatus of distribution theory.

Obviously the kinetic energy-momentum tensor $\tau_{\rm kin}^{\m\n}$ \eref{tslin} is well-defined by itself and needs no renormalization. Eventually  we define then the {\it total} renormalized energy-momentum tensor as
\be\label{tmnr}
T^{\m\n}=T^{\m\n}_f +T^{\m\n}_{int} + \tau_{\rm kin}^{\m\n},
\ee
that by construction is a distribution and coincides in the complement of the world-sheet with the original - bare - energy-momentum \eref{123}.

\subsection{Energy-momentum conservation and self-force}\label{emca}

Both properties $a)$ and $b)$ play an essential role in the implementation of energy-momentum conservation and in the derivation of the self-force. At the end of section \ref{cf} we saw that the bare field energy-momentum tensor has the property
\[
\pa_\m \tau^{\m\n}_f(x)=0, \quad\mbox{if $x$ belongs to the complement of the world-sheet,}
\]
thanks to the fact that the fields satisfy the linearized equations of motion \eref{eqfil}-\eref{eqgl}. But since  by construction - see condition b) above -  the tensor $T^{\m\n}_f$ \eref{tmnf} equals $\tau^{\m\n}_f$ in the complement of the world-sheet, it follows that the {\it distributional} divergence $\pa_\m T^{\m\n}_f$ is supported on the world-sheet. Since also $T^{\m\n}_{int}$ is supported on the world-sheet, and our whole construction preserves Lorentz- as well reparameterization-invariance, we derive the distributional relation
\be\label{ident1}
\pa_\m \left(T^{\m\n}_f + T^{\m\n}_{int}\right)=-\int {\cal S}^\n  \dl^D(x-y(\s))\sqrt{\g}\,d^2\s,
\ee
where ${\cal S}^\n$ is some covariant vector defined on the world-sheet. Since the kinetic energy-momentum tensor of the string \eref{tslin} satisfies the identity, see \eref{acclin},
\be
\pa_\m \tau_{\rm kin}^{\m\n}=  Me^{\bt\Psi}\int \Delta_i U^{\n i} \, \dl^D(x-y(\s))\sqrt{\g}\,d^2\s,
\ee
imposing on the tensor \eref{tmnr} total energy-momentum conservation we obtain
\be\label{iden2}
\pa_\m T^{\m\n}= \int \left(Me^{\bt\Psi}\Delta_i U^{\n i} - {\cal S}^\n\right) \dl^D(x-y(\s))\sqrt{\g}\,d^2\s= 0.
\ee
In this way we deduce the equation of motion for the self-interacting string
\be\label{eqself}
Me^{\bt\Psi}\Delta_i U^{\m i} = {\cal S}^\m,
\ee
replacing the ill-defined equation \eref{ill}. Equation \eref{eqself} identifies the vector ${\cal S}^\m$ showing up in \eref{ident1} as the {\it self-force}.

As anticipated in the introduction, it could happen that ${\cal S}^\m$ is not a multiplicative vector, but contains also derivative operators, like
 \be\label{sder}
 {\cal S}^\m\sim \pa^\m+A^\m{}_\n\,\pa^\n+\cdots.
\ee
In this case \eref{iden2} could no longer be made to vanish upon imposing \eref{eqself}. As we will see, even in the most simplest case of a string in uniform motion, ${\cal S}^\m$ will actually contain terms like \eref{sder}, but those terms can always be eliminated thanks to the  finite-counterterm-ambiguity.

Concerning this strategy to derive the self-force we insist on the fact that, in presence of singularities, there is no longer a {\it fundamental} principle - as the action principle - allowing to derive the dynamics of a theory, in particular the self-force.  As we observed already, the alternative strategy based on the direct renormalization of the bare self-force \eref{ill}, as done {\it e.g.} in \cite{BD12bis,BCM}, entails no control on energy-momentum conservation: if the singularities are {\it too strong}, this strategy  may even turn out to violate energy-momentum conservation, in which case it must be rejected; for a concrete example - regarding massless charges in four dimensions - see \cite{KL2}. The physical meaning of this potential conflict between different procedures to derive the dynamics of self-interacting objects in {\it extremal} cases, is still an open problem, to be investigated further. Its origin is however clear: the failure of the action principle to describe self-interactions.

\section{Strings in uniform motion}\label{siu}

In this section we apply the procedure outlined in sections \ref{rem} and \ref{emca} to a flat string moving uniformly - in which case the entire program can be carried out analytically - thereby illustrating its internal consistency in a simple, although non-trivial, physical situation. In this case we expect of course to gain ${\cal S}^\m=0$. As above, in the following we will work out the details in the Dirac framework, relegating the differences that occur in the other two frameworks to sections \ref{llf} and \ref{cfw}.

The world-sheet swept out by a string in uniform motion has the form
\be\label{sflat}
y^\m(\s)=U^\m_i\s^i
\ee
and correspondingly the tangent vectors $U^\m_i=\pa_i y^\m(\s)$ are {\it constant}, as are the geometric objects in \eref{flat}.

\subsection{Regularized fields and energy-momentum tensors}\label{rfa}

For a configuration like \eref{sflat} the regularized fields
\eref{fxe} can computed analytically, upon reading the {\it currents} $j(\s)$  from \eref{square} and \eref{eqfil}-\eref{eqgl} and inserting the regularized Green functions \eref{greene}. The integral over the $\s^i$ in \eref{fxe} can be carried out explicitly for even as well as for odd $D$ - see {\it e.g.} the appendix in reference \cite{KL1} - and the regularized fields $f_\ve$ have the same analytical form for all $D>4$:
\begin{align}\label{fe1}
\vp_\ve(x)&=\displaystyle\frac{\bt G M e^{\bt\Psi}}{(4-D)\Omega_{D-2}
(-k_{\a\bt}x^\a x^\bt+\ve^2)^{\frac{D}{2}-2}},\\
B_\ve^{\m\n}(x)&=\displaystyle\frac{G\La e^{2\a\Psi}}{(4-D)\Omega_{D-2}
(-k_{\a\bt}x^\a x^\bt+\ve^2)^{\frac{D}{2}-2}}\,w^{\m\n},\\
F_\ve^{\m\n}(x)&=\displaystyle\frac{GM e^{\bt\Psi}}{(4-D)\Omega_{D-2}
(-k_{\a\bt}x^\a x^\bt+\ve^2)^{\frac{D}{2}-2}}\,l^{\m\n},\label{fe3}
\end{align}
where we introduced the $(D-2)$-dimensional solid angle
\[
\Omega_{D-2}=\frac{2\pi^{\frac{D-2}{2}}}{\G\left(\frac{D-2}{2}\right)}.
\]
For $D=4$ the integrals \eref{fxe} are infrared divergent due to the infinite spatial extension of a flat string. This is merely an artifact of the Green-function method, that for infinitely extended strings in $D=4$ does not work properly\footnote{This is similar to the failure of the Green-function method to solve Maxwell's equations in $D=4$ in the case of a charged particle moving along a straight line at the speed of light \cite{AL2}.}. In this case it is however easy to solve the equations \eref{square} from scratch\footnote{Alternatively one may introduce an infrared cut-off $l$ for the coordinate $\s^1$ in \eref{fxe}, imposing $\s^1<l$, and send then $l\ra \infty$; equations \eref{fe14}-\eref{fe34} are then regained identifying $l\leftrightarrow \la$. Formally the expressions \eref{fe14}-\eref{fe34} could also be obtained performing in \eref{fe1}-\eref{fe3} the limit $D\ra 4$  and identifying $\frac{1}{D-4}\leftrightarrow\ln \la$.}, and regularized solutions can be obtained upon replacing $k_{\a\bt}x^\a x^\bt \ra k_{\a\bt}x^\a x^\bt-\ve^2$:
\begin{align}\label{fe14}
\vp_\ve(x)&=\displaystyle\frac{\bt G M e^{\bt\Psi}}{4\pi}
\ln\left(\frac{-k_{\a\bt}x^\a x^\bt+\ve^2}{\la^2}\right),\\
B_\ve^{\m\n}(x)&=\displaystyle\frac{G\La e^{2\a\Psi} w^{\m\n}}{4\pi}\ln\left(\frac{-k_{\a\bt}x^\a x^\bt+\ve^2}{\la^2}\right),\\
F_\ve^{\m\n}(x)&=\displaystyle\frac{GM e^{\bt\Psi}l^{\m\n}}{4\pi}
\ln\left(\frac{-k_{\a\bt}x^\a x^\bt+\ve^2}{\la^2}\right).\label{fe34}
\end{align}
For dimensional reasons we are obliged to introduce a parameter $\la$ with the dimension of length - in principle a new constant of the theory. When computing the field-strengths $\pa_\m f_\ve(x)$, appearing in the the regularized field energy-momentum tensor $\tau^{\m\n}_{f\ve}$, the constant $\la$ drops out. It will however survive in the regularized interaction energy-momentum tensor $\tau^{\m\n}_{int\,\ve}$, see \eref{tint}, where the fields \eref{fe14}-\eref{fe34} are evaluated on the world-sheet.

The regularized fields \eref{fe1}-\eref{fe34} depend in a simple way on $x^\m$ through the factor $-k_{\a\bt}x^\a x^\bt+\ve^2$, that is positive definite since the orthogonal projector to the string $k_{\a\bt}$ is negative definite. They depend in particular only on the $D-2$ orthogonal coordinates $k^{\m\n} x_\n$. The fields \eref{fe1}-\eref{fe34} are regular on the world-sheet: for $x^\m= y^\m(\s)=U^\m_i\s^i$ we in fact have $-k_{\a\bt}x^\a x^\bt+\ve^2=\ve^2\neq 0$. These fields are actually of class $C^\infty$ in whole $\mathbb{R}^D$ for all $D\ge 4$. For $\ve=0$, near the string they exhibit the singular behavior anticipated in \eref{fxdiv}.

Inserting the (derivatives of the) fields \eref{fe1}-\eref{fe34} in \eref{tf} we obtain a single analytic expression for the regularized field energy-momentum tensor, valid for all $D\ge 4$:
\be\label{tmnfe}
\tau^{\m\n}_{f\ve}=\frac{G}{\Omega^2_{D-2}(-k_{\a\bt}x^\a x^\bt+\ve^2)^{D-2}}\left(C\left(
k^{\m\a}k^{\n\bt}x_\a x_\bt-\frac{1}{2}\,\eta^{\m\n}k_{\a\bt}x^\a x^\bt\right)-\La^2 e^{2\a\Psi}l^{\m\n}k_{\a\bt}x^\a x^\bt\right).
\ee
The coefficient $C$ has the expression
\be\label{c}
C= M^2e^{2\bt\Psi}\left(\bt^2+\frac{D-4}{D-2}\right)-\La^2e^{2\a\Psi},
\ee
which in the {\it fundamental string} model \eref{fundpar} is zero for all $D\ge4$.
In \eref{tmnfe} the contributions from the scalar field are those proportional to $M^2\bt^2$, those from the gravitational field are the ones proportional to $M^2$, and the ones from the axion are proportional to $\La^2$.

For what concerns the regularized interaction energy-momentum tensor, substituting \eref{fe1}-\eref{fe3} in \eref{tint} for $D>4$ we obtain
\be\label{tinte}
\tau^{\m\n}_{int\,\ve}=\frac{GM^2e^{2\bt\Psi}}{(4-D)\Omega_{D-2}\,\ve^{D-4}}\left(\bt^2
+\frac{D-4}{D-2}\right)\int l^{\m\n}\dl^D(x-y)\sqrt{\g}\,d^2\s,
\ee
while  for $D=4$ from \eref{fe14}-\eref{fe34} we get\footnote{\label{limit}\eref{tinte4} can be obtained from \eref{tinte} considering  the limit $D\ra 4$ and  identifying $\frac{1}{D-4}\leftrightarrow\ln \la$.}
\be\label{tinte4}
\tau^{\m\n}_{int\,\ve}=\frac{GM^2\bt^2e^{2\bt\Psi}\ln(\ve/\la)}{2\pi}
\int l^{\m\n}\dl^4(x-y)\sqrt{\g}\,d^2\s.
\ee
For strings in uniform motion these tensors have thus purely a divergent part,
\[
\tau_{int\,\ve}^{\m\n}\big|_{div}= \tau_{int\,\ve}^{\m\n},
\]
so that the renormalized  interaction energy-momentum tensors \eref{tmnint} vanish for all $D\ge4$,
\be\label{tzero}
T^{\m\n}_{int}=0.
\ee
We see that the divergent counterterm $\tau_{int\,\ve}^{\m\n}\big|_{div}$ is non-vanishing for all $D\ge4$, for both our string-models: in the {\it general} model the parameters are arbitrary, and in the {\it fundamental string} model \eref{fundpar} we have $\bt^2=\frac{2}{D-2}$. Notice, however, that in $D=4$ the gravitational field - even in the {\it general} model - does not contribute to $\tau_{int\,\ve}^{\m\n}$: in \eref{tinte4} there is  in fact no term proportional to $M^2$, but only a term proportional to $\bt^2M^2$ coming from the dilaton.

As anticipated, the renormalization of the field energy-momentum tensors \eref{tmnfe} is more involved  since its support is the bulk $\mathbb{R}^D$; we face it in the next sections.

\subsection{Renormalization: an example}

As $\ve$ tends to zero point-wise in \eref{tmnfe}, we obtain a function $\tau^{\m\n}_{f}(x)$ - the bare energy-momentum tensor - that is regular for $k^{\m\n} x_\n\neq 0$, {\it i.e.} in the complement of  the world-sheet. In the vicinity of the world-sheet  $\tau^{\m\n}_{f}(x)$ behaves, however, as in \eref{tmndiv} and  is thus not a distribution. To isolate the divergent counterterm $\tau^{\m\n}_{f\ve}\big|_{div}$ of $\tau^{\m\n}_{f\ve}$, that  diverges as $\ve\ra 0$ in the distributional sense, we use a technique that we illustrate first in a simple example. The results of the actual calculation of $\tau^{\m\n}_{f\ve}\big|_{div}$, first for $D=4$ and then for $D>4$, will be given subsequently.

Consider the functions of a single variable
\[{\cal T}_\ve(x)=\frac{1}{(x^2+\ve^2)^2},
\]
depending on a positive real parameter $\ve$ with the dimension of a length. For every $\ve>0$ these functions represent distributions ${\cal T}_\ve\in {\cal S}'(\mathbb R)$.  More precisely, if we apply them to a test function $\vp\in {\cal S}(\mathbb R)$ the resulting integrals
\be\label{fef}
{\cal T}_\ve(\vp)= \int \frac{\vp(x)}{(x^2+\ve^2)^2}\,dx
\ee
are convergent\footnote{Actually, for ${\cal T}_\ve$ to be elements of  ${\cal S}'(\mathbb R)$ the quantities $|{\cal T}_\ve(\vp)|$ must be dominated by (a finite sum of) {\it semi-norms} of $\vp$.} for every $\vp$. The {\it point-wise} limit for $x\neq 0$
\be\label{point}
\lim_{\ve\ra0}{\cal T}_\ve(x)=\frac{1}{x^4}
\ee
does however not represent a distribution, because the integrals
\be
\int\frac{ \vp(x)}{x^4}\,dx
\ee
diverge due to the non-integrable singularity at $x=0$.

We want to overcome this difficulty at the price of modifying ${\cal T}_\ve(x)$ as little as possible, {\it i.e.} only at $x=0$. To this order we isolate the singularity at $x=0$ in \eref{fef} writing
\ba\label{fef1}\nn
{\cal T}_\ve(\vp)&=& \int \frac{\vp(x)-\vp(0)-\frac{x^2}{2}\,\vp''(0)}{(x^2+\ve^2)^2}\,dx+
\vp(0) \int \frac{dx}{(x^2+\ve^2)^2}+\frac{1}{2}\,\vp''(0) \int \frac{x^2dx}{(x^2+\ve^2)^2}\\
&= &\int \frac{\vp(x)-\vp(0)-\frac{x^2}{2}\,\vp''(0)}{(x^2+\ve^2)^2}\,dx + \frac{\pi}{2\ve^3}\,\vp(0)+\frac{\pi}{4\ve}\,\vp''(0).\label{exam}
\ea
Since the first integral in \eref{exam} converges now as $\ve\ra0$ for every $\vp\in {\cal S}(\mathbb R)$, we read off the ``divergent part'' of ${\cal T}_\ve$ as
\be\label{divf}
{\cal T}_\ve\big|_{div}= \frac{1}{(x^2+\ve^2)^2}\,\bigg|_{div}= \frac{\pi}{2\ve^3}\,\dl(x)+\frac{\pi}{4\ve}\,\dl''(x).
\ee
${\cal T}_\ve\big|_{div}$ contains a leading divergence, supported in $x=0$, proportional to $1/\ve^3$, and a sub-leading one - yet supported in $x=0$ - proportional to $1/\ve$: the general lesson is that the stronger the divergences (higher inverse powers of $x$) present in ${\cal T}_\ve$, the more terms proportional to higher derivatives of the $\dl$-function (higher inverse powers of $\ve$) are present in ${\cal T}_\ve\big|_{div}$.

Subtracting the  ``divergent counterterm'' we conclude then that the distributional limit
\be\label{renf}
{\cal S}'-\lim_{\ve\ra0}\left({\cal T}_\ve- {\cal T}_\ve\big|_{div}\right)
\equiv {\cal T}
\ee
exists and defines the renormalized  version of the function \eref{point}. The explicit expression of ${\cal T}$ is
\[
{\cal T}(\vp)=\int \frac{\vp(x)-\vp(0)-\frac{x^2}{2}\,\vp''(0)}{x^4}\,dx.
\]
We have thus achieved our goal: from \eref{exam}, \eref{divf} and \eref{renf} we  deduce that ${\cal T}$ is a {\it distribution}, that in $\mathbb{R}\setminus\{0\}$  coincides with $1/x^4$, {\it i.e.} with the point-wise limit \eref{point}\footnote{The precise meaning of this is that when applied to a test function $\vp(x)$ that vanishes in an arbitrarily small neighborhood of $x=0$, the function $1/x^4$ and the distribution ${\cal T}$ give the same value. In the case at hand ${\cal T}$ could actually be written as the distributional derivative of a basic distribution, {\it i.e.} of the {\it principal part} of $1/x$, namely ${\cal T}=-\frac{1}{6}(d/dx)^3P(1/x)$.}.

\subsubsection{Subtraction schemes and finite counterterms}

In choosing the divergent part \eref{divf} we tacitly ``resolved'' an indeterminacy regarding the {\it finite} part of ${\cal T}$ - relying on what in quantum field theory would be called a {\it minimal subtraction scheme}. In fact, the ``renormalized'' distribution ${\cal T}$ is determined only modulo the {\it finite local counterterms}
\[
{\cal T}\ra {\cal T}+a\,\dl(x)+b\,\dl''(x),
\]
where we omitted odd derivatives of the $\dl$-function to preserve the invariance under parity of ${\cal T}_\ve$. In the present case the choice \eref{divf} might be justified because the coefficients $a$ and $b$ must be dimensionful, {\it i.e.} of length dimension respectively $1/L^3$ and $1/L$. If no fundamental constants with inverse length-dimensions show up in the theory, then $a$ and $b$ must actually vanish.

Consider with this respect the further example
\[
{\cal U}_\ve(x)=\frac{1}{|x|+\ve},
\]
whose divergent part is
\[
{\cal U}_\ve\big|_{div}=-2\ln(\ve/L)\,\dl(x).
\]
In this case, for dimensional reasons the separation of the divergent part required the introduction of an arbitrary parameter $L$ with the dimension of a length. This leads in the renormalized distribution
\[
{\cal U} ={\cal S}'-\lim_{\ve\ra0}\big({\cal U}_\ve+2\ln(\ve/L)\,\dl(x)\big),
\]
to an indeterminacy of the type
\[
{\cal U}\ra {\cal U}+a\,\dl(x),
\]
where $a$ is a dimensionless parameter, that {\it a priori} can not be set to zero. This is a simple example of the {\it finite-counterterm-ambiguity}, that will play a significant role in sections \ref{fc} and \ref{ufc}.

\subsection{Renormalization of the field energy-momentum tensor in $D=4$}

The determination of the divergent counterterm of the tensors \eref{tmnfe} relies on a straightforward generalization of the above example. Due to its obvious relevance we analyze first the four-dimensional case.

For $D=4$ \eref{tmnfe} reduces to
\be\label{tmnfe4}
\tau^{\m\n}_{f\ve}=\frac{G}{4\pi^2(-k_{\a\bt}x^\a x^\bt+\ve^2)^2}\left(C_4\left(
k^{\m\a}k^{\n\bt}x_\a x_\bt-\frac{1}{2}\,\eta^{\m\n}k_{\a\bt}x^\a x^\bt\right)-\La^2 e^{2\a\Psi}l^{\m\n}k_{\a\bt}x^\a x^\bt\right),
\ee
where
\[
C_4=M^2\bt^2e^{2\bt\Psi}-\La^2e^{2\a\Psi}.
\]
The formula analogous to \eref{divf} we need is
\be\label{form4}
{k^{\m\a}\,k^{\n\bt}\,x_\a x_\bt \over \left(-k_{\a\bt}\,x^\a x^\bt+\ve^2\right)^2}\,\bigg|_{div}= \pi\ln (\ve/L) \int \left(\eta^{\m\n}-l^{\m\n}\right)\dl^4(x-y)\sqrt{\g}\,d^2\s.
\ee
There is only a logarithmic divergence, since near the world-sheet for $\ve=0$ the l.h.s. of \eref{form4}
diverges as $x_\perp^2$, and the orthogonal space is tow-dimensional. For dimensional reasons we are obliged to introduce an arbitrary length scale $L$, that reflects the subtraction-scheme ambiguity discussed above.

Applying \eref{form4} to \eref{tmnfe4} we obtain the divergent counterterm
\be\label{div4}
\tau^{\m\n}_{f\ve}\big|_{div}=-\frac{G\ln(\ve/L)}{4\pi}\left(M^2\bt^2e^{2\bt\Psi}+\La^2 e^{2\a\Psi}\right) \int l^{\m\n}\dl^4(x-y)\sqrt{\g}\,d^2\s.
\ee
As in the case of the interaction energy-momentum tensor \eref{tinte4}, also in \eref{div4} there is no divergent contribution from the gravitational field.
Given \eref{tmnfe4} and \eref{div4}, the distributional limit
\be\label{tmnf4}
T^{\m\n}_f\equiv{\cal S}'-\lim_{\ve\ra0}\left(\tau_{f\ve}^{\m\n}-\tau_{f\ve}^{\m\n}\big|_{div}\right)
\ee
exists now and defines the renormalized field energy-momentum tensor.

\subsubsection{Cancelation of divergences}\label{nrt}

Within our approach the energy-momentum tensors are always ``renormalizable'' - in the sense that the divergent counterterms are localized on the world-sheet - so that the vanishing of the divergent counterterms is actually not of central importance. Nevertheless it is instructive, also for the comparison with known results in the literature, to see if there are models for which the divergences cancel out. To make this analysis comparative we anticipate some results from later sections.

The divergent counterterms \eref{tinte4} and \eref{div4} are non-vanishing in the {\it general} model as well as in the {\it fundamental string} model, unless $\La=\bt=0$. The situation is different for what concerns the {\it total} counterterm (we omit finite terms)
\be
 \tau^{\m\n}_{int\,\ve}\big|_{div} + \tau^{\m\n}_{f\ve}\big|_{div} =
 \frac{G}{4\pi} \,\ln(\ve/L) \left(M^2\bt^2e^{2\bt\Psi}-\La^2e^{2\a\Psi}\right)\int l^{\m\n}\dl^4(x-y)\sqrt{\g}\,d^2\s.
\label{tot4}
\ee
For the {\it general} model this is still divergent, while for the {\it fundamental string} model \eref{fundpar} the divergences actually cancel. The cancelation occurs between the dilaton  $(M^2\bt^2)$ and the axion $(\La^2)$, while, as we observed above, the divergences of the gravitational field $(M^2)$ just drop out, even in the {\it general} model. This result proves in particular, for $D=4$, the compensation between field-divergences, originating from the bulk, and  interaction-divergences, genuinely localized on the world-sheet, conjectured in the effective-action approach \cite{BD12}. As we will see in section \ref{rot}, for $D>4$ this compensation will no longer occur, neither in the Dirac framework that we are applying here, nor in the Landau-Lifshitz and canonical frameworks. Nevertheless in the last two frameworks the field-divergences and  interaction-divergences will cancel {\it separately} for all $D\ge4$, see sections~\ref{llf} and \ref{cfw}.

The cancelation of {\it gravitational} divergences, noticed previously, is special to $D=4$ and occurs - even in the {\it general} model - {\it separately} in $\tau^{\m\n}_{int\,\ve}|_{div}$ \eref{tinte4} and  $\tau^{\m\n}_{f\ve}|_{div}$ \eref{div4}. This separate cancelation occurs in the Dirac framework and, as we will see, in the canonical framework, while in the Landau-Lifshitz framework the gravitational field-divergences will cancel against the gravitational interaction-divergences. In general the pattern of cancelation of divergences, even in $D=4$, is thus {\it framework-dependent}.

A characteristic feature of the four-dimensional total counterterm \eref{tot4} is that, being proportional to $l^{\m\n}$, it {\it could} be eliminated via the string-tension redefinition
(see \eref{tslin})
\be\label{redef}
M\rightarrow M'= M +  \frac{G}{4\pi} \left(M^2\bt^2e^{\bt\Psi}-\La^2e^{(2\a-\bt)\Psi}\right)\ln(\ve/L).
\ee
In contrast, in dimensions $D>4$ there will be several different tensorial structures showing up in the divergent counterterms, whose cancelation could not be achieved renormalizing the parameters of the original theory: in the general case, {\it by-hand} subtractions of divergences, as in \eref{tmnf4}, represent thus a basic ingredient of our approach.

From the presence of the $(\ln L)$-term in \eref{tot4} we conclude that in $D=4$  the finite-counterterm-ambiguity amounts simply to a redefinition of the string tension.

\subsection{Renormalization of the field energy-momentum tensor in $D>4$}\label{rot}

To determine the divergent counterterm of the field energy-momentum tensor \eref{tmnfe} in a generic space-time, we need the generalization of \eref{form4} to a generic $D\ge4$ (see reference~\cite{KL1})
\be\label{formD}
{k^{\m\a}\,k^{\n\bt}\,x_\a x_\bt \over \left(-k_{\a\bt}\,x^\a x^\bt+\ve^2\right)^{D-2}}\,\bigg|_{div}=
\sum_{j=0}^{D-4}{}'A_j\int\big((l^{\m\n}-\eta^{\m\n})\,\square-j\,
\pa^\m\pa^\n\big)\Box^{j/2-1}\,\dl^D(x-y)\sqrt{\g}\,d^2\s.
\ee
The {\it prime} indicates that the sum extends only over {\it even} $j$ and the coefficients $A_j$ (divergent for $\ve\ra0$) are given by
 \be
A_j=\begin{cases}
\displaystyle {(-)^{j/2}\,\pi^{\frac{D-2}{2}}\,\G\left({D-j\over 2}-2\right)\over 2^{j+1}\,\G(D-2)\,\G\left({j\over 2}+1\right)}\cdot{1\over \ve^{D-j-4}}, & \mbox{for }j< D-4, \\
& \\ \label{boh}
\displaystyle {(-\pi)^{\frac{D-2}{2}}\over 2^{D-4}\,\G(D-2)\G\left({D\over2}-1\right)}\cdot\ln \left(\ve/L\right), & \mbox{for } j=D-4.
\end{cases}
\ee
Applying \eref{formD} to \eref{tmnfe} we obtain for its divergent counterterm the expression, valid for all $D\ge4$,
\begin{align}
\tau^{\m\n}_{f\ve}\big|_{div}=\frac{G}{\Omega_{D-2}^2}\sum_{j=0}^{D-4}{}'A_j\int&\left(
\big(C+(D+j-2)\La^2e^{2\a\Psi}\big)l^{\m\n}\,\square+\frac{1}{2}\,C\big(D+j-4\big)\eta^{\m\n}
\,\square\right.
\nn\\
&-j\,C\pa^\m\pa^\n\bigg)\square^{j/2-1}\,\dl^D(x-y)\sqrt{\g}\,d^2\s.\label{divD}
\end{align}
Contrary to the four-dimensional case, this counterterm exhibits a sum of derivatives of $\dl$-functions $\pa^j\dl^D(x-y)$, multiplied by the divergent factor $1/\ve^{D-j-4}$. The leading divergence is $1/\ve^{D-4}$ and corresponds to $j=0$. The terms with $j>0$ represent an entire series of {\it subleading} divergences - absent in $D=4$ - and none of them could be eliminated through the redefinition of the tension, or some other  coupling constants, like in \eref{redef}. Notice also the appearance of {\it gravitational} divergences, {\it i.e.} the terms proportional to $(D-4)M^2$ in the coefficient $C$ in \eref{c}, that were absent in $D=4$.

Let us analyze more closer the leading divergence in \eref{divD}, that has the form
\be
\begin{aligned}
\tau^{\m\n}_{f\ve}\big|_{div}^{lead}=
{G\pi^{1/2}\G\left(\frac{D-4}{2}\right)\over 2^{D-1}\,\Omega_{D-2}\G\left(\frac{D-1}{2}\right)\ve^{D-4}}
&\int \bigg(\big(C+(D-2)\La^2e^{2\a\Psi}\big)l^{\m\n}\\
&\left.
+\frac{1}{2}\,C\big(D-4\big)\eta^{\m\n}
\right)
\dl^D(x-y)\sqrt{\g}\,d^2\s.\label{divDlead}
\end{aligned}
\ee
For $D>4$ it contains hence the two tensorial structures $l^{\m\n}$ and $\eta^{\m\n}$. Correspondingly, for $D>4$ in the {\it general} model there is no way to cancel even this leading field-divergence against the interaction-divergence \eref{tinte}, which contains only $l^{\m\n}$.

But even in the {\it fundamental string} model, where $C=0$ and the tensor $\eta^{\m\n}$ drops out, the numerical coefficients of $l^{\m\n}$ in \eref{tinte} and \eref{divDlead} do not match. We conclude thus that in the Dirac framework in the {\it fundamental string} model the total divergences cancel in $D=4$, but not for $D>4$. In particular in this framework for $D>4$ the string tension suffers a non-vanishing renormalization - a feature that would not be expected at the basis of the non-renormalization theorems of superstring amplitudes \cite{DH}. This occurrence may disfavor the Dirac-framework w.r.t. the other two frameworks, although we were not able to find a physical reason for this.

In the {\it general} model we define the renormalized field energy-momentum tensor as in \eref{tmnf4}, with $\tau^{\m\n}_{f\ve}$ and  $\tau^{\m\n}_{f\ve}\big|_{div}$ given respectively in \eref{tmnfe} and \eref{divD},
\be\label{tmnfD}
T^{\m\n}_{f(0)}\equiv{\cal S}'-\lim_{\ve\ra0}\left(\tau_{f\ve}^{\m\n}-\tau_{f\ve}^{\m\n}\big|_{div}\right).
\ee
We put a $(0)$ in the definition of $ T^{\m\n}_f$, due to the finite-counterterm-ambiguity that we will encounter in the next section.

We emphasize that the tensor \eref{tmnfD} is not a merely abstract object in that, being regular in whole space, {\it i.e.} being a distribution, it can be used to compute concretely the finite energy and momenta in arbitrary finite volumes - even if these volumes intersect the world-sheet.  $T^{\m\n}_{f(0)}$ shares this property with the renormalized  energy-momentum tensor of the electromagnetic field of a charged point-particle in four dimensions, whose integrals over a volume {\it enclosing} the particle always converge, giving rise to finite four-momenta \cite{LM}.

\subsection{Energy-momentum conservation}\label{emc}

By construction \eref{tmnfD} is a distribution and so its divergence $\pa_\m T^{\m\n}_{f(0)}$ is perfectly well-defined. From the general analysis of section \ref{rem} we know furthermore that  $\pa_\m T^{\m\n}_{f(0)}$ is supported on the word-sheet. To evaluate it explicitly
we use that derivatives are {\it continuous} operations in distribution space, so that in \eref{tmnfD} we can freely interchange the derivatives with the limit. Moreover, since $\tau_{f\ve}^{\m\n}$ is a regular distribution, its derivatives can be computed in the usual way.  From  \eref{tmnfe} and \eref{divD} we get
\be
\begin{aligned}
\pa_\m\left(\tau_{f\ve}^{\m\n}-\tau_{f\ve}^{\m\n}\big|_{div}\right)
=\frac{GC}{2\Omega^2_{D-2}}\,\pa^\n&\left(\frac{\ve^2}{(-k_{\a\bt}x^\a x^\bt+\ve^2)^{D-2}}\right.\\
&-
\sum_{j=0}^{D-4}{}'(D-j-4)A_j\int\square^{j/2}\,\dl^D(x-y)\sqrt{\g}\,d^2\s
\bigg),\label{cons1}
\end{aligned}
\ee
where, for convenience, we factorized out a derivative. The first term between parentheses at the right hand side, coming from the divergence of $\tau_{f\ve}^{\m\n}$, multiplies a factor of $\ve^2$. This means that, when taking $\ve\ra0$, this term is entirely supported on the world-sheet, as it must be. Applying this term to a test function $\vp$ and performing the expansion in powers of $\ve$, one gets \cite{KL1}
\be\label{factor}
\frac{\ve^2}{(-k_{\a\bt}x^\a x^\bt+\ve^2)^{D-2}}=
\sum_{j=0}^{D-4}{}'B_j\,\Box^{j/2}\,\dl^D(x-y)\sqrt{\g}\,d^2\s+o(\ve),
\ee
where
\be
\label{bj}
B_j=\begin{cases}(D-j-4)A_j,&\mbox{for }j<D-4,\\
\dis \frac{(-1)^{D/2}\Omega_{D-2}}{2^{D-3}\,\Gamma(D-2)},&\mbox{for }j=D-4,\quad (\mbox{if } D \mbox{ is even)},
\end{cases}
\ee
with $A_j$ given in \eref{boh}. In \eref{factor} with $o(\ve)$  we understood terms that converge to zero as $\ve\ra0$ in the distributional sense. We see that all divergences in \eref{cons1} cancel out, as they must by construction. However, for $D$ {\it even} the expansion \eref{factor} contains also a non-vanishing {\it finite} term, the one corresponding to $j=D-4$. Consequently, for the divergence of the energy-momentum tensor \eref{tmnfD} we get
\begin{numcases}
{\pa_\m T^{\m\n}_{f(0)}=}
0, & for $D$ odd,\label{odd}\\
\dis\frac{GCB_{D-4}}{2\Omega_{D-2}^2} \int\pa^\n\,\square^{\frac{D-4}{2}}\,\dl^D(x-y)\sqrt{\g}\,d^2\s,&for $D$  even.\label{even}
\end{numcases}
For the four-dimensional string we have for example
\[
\pa_\m T^{\m\n}_{f(0)}=\frac{GC}{8\pi} \int\pa^\n\dl^4(x-y)\sqrt{\g}\,d^2\s.
\]

\subsubsection{Finite counterterms}\label{fc}

In principle, according to our approach the {\it anomaly} encountered in \eref{even} for $D$ even - a non-vanishing $D$-divergence for the otherwise well-behaved distribution  $T^{\m\n}_{f(0)}$ - determines the self-force ${\cal S}^\m$. Recalling that the renormalized interaction energy-momentum tensor \eref{tzero} is zero, from \eref{ident1} and \eref{even} we would then get an ${\cal S}^\m$ that is a derivative operator, and not a simply a vector. There would thus exist no string-equation of motion ensuring total energy-momentum conservation.

On the other hand it is a basic fact in any renormalization process, in quantum as well as in classical theory, that once we subtract divergent terms from a physical quantity, this quantity remains by itself determined only modulo finite terms of the {\it same structure} as the divergent ones. This offers a way out thanks to the fact that the  anomaly in \eref{even} is a {\it trivial} anomaly, in that in can - and must  - be eliminated by subtracting a finite counterterm, in very much the same way as one eliminates trivial anomalies in quantum field theory, once one has introduced a regularization that breaks a classical symmetry.

In the present case the appropriate finite counterterm is
\be\label{finc}
T^{\m\n}\big|_{fin}=\frac{GCB_{D-4}}{2\Omega_{D-2}^2} \int \eta^{\m\n} \,\square^{\frac{D-4}{2}}\,\dl^D(x-y)\sqrt{\g}\,d^2\s,
\ee
which in $D = 4$ becomes
\be\label{finitec4}
T^{\m\n}\big|_{fin}=\frac{GC}{8\pi} \int \eta^{\m\n}\dl^4(x-y)\sqrt{\g}\,d^2\s.
\ee
The final renormalized field energy-momentum tensor
\be\label{finitec}
T^{\m\n}_f= T^{\m\n}_{f(0)}-T^{\m\n}\big|_{fin}
\ee
satisfies in turn
\be
\pa_\m T^{\m\n}_f=0.
\ee
Together with \eref{tzero} equation \eref{ident1} gives then rise to a vanishing self-force, ${\cal S}^\m=0$, as is of course in line with our string moving freely in space-time.

\section{Landau-Lifshitz framework}\label{llf}

In this section we display the main changes that arise w.r.t. the preceding analysis, when we use for the gravitational field the Landau-Lifshitz pseudo-tensor $\w\Sigma^{\m\n}$  (\eref{ll} and \eref{z}) in place of the Dirac pseudo-tensor $\Sigma^\m{}_\n$ \eref{TD}.
As we saw, the use of $\w\Sigma^{\m\n}$ instead of $\Sigma^\m{}_\n$ induces in the field and interaction energy-momentum tensors the modifications \eref{fdll} and \eref{intll}, so that it is easy to extract from those relations and our previous results \eref{tmnfe} and \eref{tinte}, using still \eref{fe1}-\eref{fe34}, the new regularized tensors for a generic $D\ge4$\footnote{It is understood that the expression of $\w\tau^{\m\n}_{int\,\ve}$ for $D=4$ is obtained from \eref{tintell}, taking the appropriate limit, see footnote \ref{limit} in section \ref{rfa}.}
\begin{align}
\label{tmnfell}
\w\tau^{\m\n}_{f\ve}&=\frac{G}{\Omega^2_{D-2}(-k_{\a\bt}x^\a x^\bt+\ve^2)^{D-2}}\left(C\left(
k^{\m\a}k^{\n\bt}x_\a x_\bt-\frac{1}{2}\,\eta^{\m\n}k_{\a\bt}x^\a x^\bt\right)+K l^{\m\n}k_{\a\bt}x^\a x^\bt\right),\\
\label{tintell}
\w\tau^{\m\n}_{int\,\ve}&=\frac{GM^2e^{2\bt\Psi}N}{(4-D)\Omega_{D-2}\,\ve^{D-4}}\int l^{\m\n}\dl^D(x-y)\sqrt{\g}\,d^2\s= \w\tau^{\m\n}_{int\,\ve}\big|_{div}.
\end{align}
We introduced the coefficients ($C$ is the same as in \eref{c})
\begin{align}\label{c1}
C&=\dis M^2e^{2\bt\Psi}\left(\bt^2+\frac{D-4}{D-2}\right)-\La^2e^{2\a\Psi},\\
K&=M^2e^{2\bt\Psi}-\La^2 e^{2\a\Psi},\label{k1}\\
N&=\bt^2-\frac{2}{D-2}\label{n1}.
\end{align}
Notice that w.r.t. \eref{tmnfe} in \eref{tmnfell} only the coefficient of the last term changed. The divergent counterterm of \eref{tmnfell} has correspondingly a structure very similar to \eref{divD}
\be
\begin{aligned}
\w\tau^{\m\n}_{f\ve}\big|_{div}=\frac{G}{\Omega_{D-2}^2}\sum_{j=0}^{D-4}{}'A_j\int&\left(
\big(C-(D+j-2)K\big)l^{\m\n}\,\square+\frac{1}{2}\,C\big(D+j-4\big)\eta^{\m\n}
\,\square\right.
\\
&-j\,C\pa^\m\pa^\n\bigg)\square^{j/2-1}\,\dl^D(x-y)\sqrt{\g}\,d^2\s.\label{divDll}
\end{aligned}
\ee

\subsection{$D=4$}

Specializing the above formulae to $D=4$ we obtain
\begin{align}
\label{div4ll}
\w\tau^{\m\n}_{f\ve}\big|_{div}&=-\frac{G\ln(\ve/L)}{4\pi}\left((\bt^2-2)M^2
e^{2\bt\Psi}+\La^2 e^{2\a\Psi}\right) \int l^{\m\n}\dl^4(x-y)\sqrt{\g}\,d^2\s,
\\
\label{tinte4ll}
\w\tau^{\m\n}_{int\,\ve}\big|_{div}&=\frac{G\ln(\ve/\la)}{2\pi}\,(\bt^2-1)M^2e^{2\bt\Psi}
\int l^{\m\n}\dl^4(x-y)\sqrt{\g}\,d^2\s.
\end{align}
For the total counterterm, disregarding finite terms, we get then
\be\label{tot4ll}
 \w\tau^{\m\n}_{int\,\ve}\big|_{div} + \w\tau^{\m\n}_{f\ve}\big|_{div} =
 \frac{G}{4\pi}\ln(\ve/L)
\left(M^2\bt^2e^{2\bt\Psi}-\La^2e^{2\a\Psi}\right)\int l^{\m\n}\dl^4(x-y)\sqrt{\g}\,d^2\s.
\ee
Comparing with the Dirac-framework results we notice first of all that the total divergent counterterm \eref{tot4ll} matches exactly  \eref{tot4}. The main difference is, however, the appearance of {\it gravitational} divergences in \eref{div4ll} as well as in \eref{tinte4ll},  proportional respectively to $M^2e^{2\bt\Psi}/2\pi$ and $-M^2e^{2\bt\Psi}/2\pi$, which are absent in \eref{div4} and \eref{tinte4}. In the sum \eref{tot4ll} they cancel therefore out.

In the {\it general} model there are again no cancelations, while in the {\it fundamental string} model \eref{fundpar}  - a further main difference w.r.t. the Dirac framework - the field-divergences and the interaction-divergences cancel now {\it separately}
\be\label{vanis}
\w\tau^{\m\n}_{f\ve}\big|_{div}=0= \w\tau^{\m\n}_{int\,\ve}\big|_{div}.
\ee
These results support in particular the hypothesis formulated in \cite{BD12} to explain the apparently contradictory results of the analysis of \cite{DH,CHH}, concerned with
field-energy-divergencies of static strings in $D=4$. The authors of \cite{DH,CHH}
found indeed that the total field-divergences cancel, whilst the gravitational field-divergences alone did not. Since the authors of \cite{BD12} - within their effective-action approach - found that in $D=4$ there were no divergent gravitational divergences contributing to the tension renormalization, they hypothesized that the gravitational field-divergences revealed in \cite{DH,CHH} should cancel against gravitational interaction-divergences. Yet the total divergences had to cancel.  All these expectations are precisely confirmed by our formulae \eref{div4ll}-\eref{vanis}\footnote{Actually the authors of \cite{DH,CHH} do not specify which gravitational energy-momentum pseudo-tensor they use. To be precise, what we have shown above is that the Landau-Lifshitz choice is consistent with their results.}.

\subsection{$D>4$}

Coming back to generic dimensions $D>4$, we notice that in the {\it fundamental string} model the coefficients $C$, $K$ and $N$ in \eref{c1}-\eref{n1} are all zero. Given \eref{tintell} and \eref{divDll} this implies that in this model the identities \eref{vanis} hold for all dimensions $D\ge4$, meaning that all {\it leading and subleading} field-divergences and  interaction-divergences cancel separately. However, for $D>4$ there is no compensation between these two types of divergences: in particular $\w\tau^{\m\n}_{int\,\ve}\big|_{div}$ \eref{tintell} contains only the leading divergence $1/\ve^{D-4}$, while  $\w\tau^{\m\n}_{f\ve}\big|_{div}$ \eref{divDll} contains also the subleading divergences $1/\ve^{D-4-j}$ for all even $0<j\le D-4$.

In the {\it fundamental string} model it happens actually that the regularized tensors $\w\tau^{\m\n}_{f\ve}$ \eref{tmnfell} and $\w\tau^{\m\n}_{int\,\ve}$ \eref{tintell}
vanish identically: this feature is characteristic for strings in uniform motion, while for accelerated strings these tensors will obviously be different from zero, see section~\ref{tgc}.

In the {\it general} model the divergent counterterms are non-vanishing and must be subtracted, as in \eref{tmnfD}. Since w.r.t. the Dirac framework the divergent counterterms changed only by terms proportional to $l^{\m\n}$ - see  \eref{tinte} {\it versus} \eref{tintell}  and \eref{divD} {\it versus} \eref{divDll} - the (distributional limit of the) divergence $\pa_\m\left(\w\tau_{f\ve}^{\m\n}-\w\tau_{f\ve}^{\m\n}\big|_{div}\right)$ is the same as in the Dirac framework. This implies that also the finite counterterm \eref{finc} to be subtracted remains the same. The renormalized field energy-momentum tensor in the Landau-Lifshitz framework is therefore
\be\label{tfinll}
\w T^{\m\n}_f ={\cal S}'-\lim_{\ve\ra0} \left(\w\tau_{f\ve}^{\m\n}-\w\tau_{f\ve}^{\m\n}\big|_{div}\right)   - T^{\m\n}\big|_{fin},\quad\quad  \pa_\m\w T^{\m\n}_f=0.
\ee
Similarly the renormalized interaction energy-momentum tensor is again zero, $\w T^{\m\n}_{int} ={\cal S}'-\lim_{\ve\ra0} \left(\w\tau_{int\,\ve}^{\m\n}-\w\tau_{int\,\ve}^{\m\n}\big|_{div}\right)=0$, as is the self-force.

\section{Canonical framework}\label{cfw}

From \eref{fdcan} and \eref{intcan} - proceeding as above - in the canonical framework we obtain
\begin{align}
\label{tmnfec}
\wh\tau^{\m\n}_{f\ve}&=\frac{GC}{\Omega^2_{D-2}(-k_{\a\bt}x^\a x^\bt+\ve^2)^{D-2}}\left(
k^{\m\a}k^{\n\bt}x_\a x_\bt-\frac{1}{2}\,\eta^{\m\n}k_{\a\bt}x^\a x^\bt\right),\\
\label{tintec}
\wh\tau^{\m\n}_{int\,\ve}&=\frac{GC}{(4-D)\Omega_{D-2}\,\ve^{D-4}}\int l^{\m\n}\dl^D(x-y)\sqrt{\g}\,d^2\s= \wh\tau^{\m\n}_{int\,\ve}\big|_{div}.
\end{align}
Contrary to the Dirac and Landau-Lifshitz frameworks, in the canonical framework the axion
contributes now also to the interaction tensor $\wh \tau_{int\,\ve}^{\m\n}$.

The divergent counterterm of the field energy-momentum tensor becomes now
\be
\wh\tau^{\m\n}_{f\ve}\big|_{div}=\frac{GC}{\Omega_{D-2}^2}\sum_{j=0}^{D-4}{}'A_j
\int\left(l^{\m\n}\,\square+\frac{1}{2}\,\big(D+j-4\big)\eta^{\m\n}
\,\square-j\,\pa^\m\pa^\n\right)\square^{j/2-1}\,\dl^D(x-y)\sqrt{\g}\,d^2\s.\label{divDc}
\ee
The expressions \eref{tintec}, \eref{divDc} of the counterterms are simpler than the corresponding expressions \eref{tinte}, \eref{divD} and \eref{tintell}, \eref{divDll} of the other two frameworks. In particular the string coupling constants enter only through the single constant $C$ \eref{c} which, we recall, vanishes in the {\it fundamental string} model. In this model we have therefore for all $D\ge4$
\be\label{cancall}
\w\tau^{\m\n}_{f\ve}\big|_{div}=0= \w\tau^{\m\n}_{int\,\ve}\big|_{div},
\ee
as in the Landau-Lifshitz framework.

In a certain sense the canonical framework ``maximizes'' the cancelation of ultraviolet divergences in the {\it fundamental string} model: for all $D\ge4$ the field- and interaction-divergences cancel separately - as in the Landau-Lifshitz framework - and in $D=4$, in addition, the gravitational field-divergences and interaction-divergences cancel separately - as in the Dirac framework.

In $D=4$ we obtain in particular ($C=M^2\bt^2e^{2\bt\Psi}-\La^2e^{2\a\Psi}$)
\be\label{d4can}
\wh\tau^{\m\n}_{int\,\ve}\big|_{div} + \wh\tau^{\m\n}_{f\ve}\big|_{div}
= \frac{GC}{4\pi}
\ln(\ve/L)\int l^{\m\n}\dl^4(x-y)\sqrt{\g}\,d^2\s,\quad\quad \wh\tau^{\m\n}_{f\ve}\big|_{div} =-\frac{1}{2}\, \wh\tau^{\m\n}_{int\,\ve}\big|_{div},
\ee
so that the total counterterm coincides with the expressions \eref{tot4} and \eref{tot4ll} of the other two frameworks. For $D=4$ the {\it total} ultraviolet divergence appears thus to have {\it universal} character, in that it is framework-independent. We did not found an {\it a priori} reason for this ``coincidence'' - which does not occur for $D>4$.

For future reference we write out \eref{divDc} for $D=5$
\be
\label{divDc5}
\wh\tau^{\m\n}_{f\ve}\big|_{div}=\frac{GC}{64}
\int \frac{1}{\ve}\left(\frac{1}{2}\,\eta^{\m\n}+l^{\m\n}\right)\dl^5(x-y)\sqrt{\g}\,d^2\s,
\ee
as well as for $D=6$
\be
\wh\tau^{\m\n}_{f\ve}\big|_{div}=\frac{GC}{48\pi^2}
\int\left(\frac{1}{\ve^2}\left(\eta^{\m\n}+l^{\m\n}\right) +\ln(\ve/L)\left(
\left(\eta^{\m\n}+\frac{1}{2}\,l^{\m\n}\right)\square-\pa^\m\pa^\n\right)\right)
\dl^6(x-y)\sqrt{\g}\,d^2\s.\label{divDc6}
\ee

A part from the the simplifications showing up in formulae \eref{tmnfec}-\eref{divDc}, in the {\it general} model the divergent counterterms
must again be subtracted, and the renormalized field energy-momentum tensor $\wh T^{\m\n}_f$ is defined exactly in the same way as in \eref{tfinll}, with the same finite counterterm \eref{finc}; it satisfies still $\pa_\m\wh T^{\m\n}_f=0$. Also in the canonical framework we have of course $\wh T^{\m\n}_{int} ={\cal S}'-\lim_{\ve\ra0} \left(\wh\tau_{int\,\ve}^{\m\n}-\wh\tau_{int\,\ve}^{\m\n}\big|_{div}\right)=0$, so that the self-force vanishes, as in the other frameworks.

In the {\it general} model, by construction the renormalized field energy-momentum tensors of the three frameworks $T^{\m\n}_f$ \eref{finitec}, $\w T^{\m\n}_f$ \eref{tfinll} and $\wh T^{\m\n}_f$ - being all  divergence-less distributions - differ from each other by the distributional divergence $\pa_\rho C^{\rho\m\n}$ of an antisymmetric tensor: this means hat for strings in uniform motion these frameworks are physically equivalent.

\subsection{Comparison with the effective-action approach}

It seems not straightforward to establish a direct link between the non-renormalization property \eref{cancall} - holding in the {\it fundamental string} model where $C=0$ - and the results of the effective-action method of \cite{BD12}, applied to the same model. The latter tests indeed different physical properties w.r.t. our approach, {\it i.e.} the ultraviolet renormalization of the string tension through a computation of the (divergent) coefficient of the kinetic action $\int \! \sqrt{\g}\,d^2\s$. This computation amounts essentially to the (gaussian) functional integral over the fields of the linearized form of the action \eref{fands}, giving rise to the ``effective action''. The latter is a non-local functional of only the string coordinates $y^\m(\s)$, that contains as divergent part a term like  $M_{div}\int \!\sqrt{\g}\,d^2\s$, where $M_{div}$ is a divergent coefficient. The authors of \cite{BD12} found the proportionality relation
\be\label{mdiv}
M_{div} \propto C,
\ee
where $C$ is precisely the coefficient \eref{c1}. If we identify the effective action with the {\it total energy integrated over time} -  although it is not clear, at least to us, whether this is the correct physical interpretation of the effective action - we may compute the (divergent part of the) former integrating the $00$ components of the canonical-framework expressions \eref{tintec} and \eref{divDc} over whole space-time: the outcome is clearly $C \int \!\sqrt{\g}\,d^2\s $, times a divergent factor, in agreement with \eref{mdiv}. In this sense the effective-action approach appears to parallel the {\it canonical} framework, while in the other two frameworks the total energy integrated over time produces a divergent coefficient in front of $\int \!\sqrt{\g}\,d^2\s$, that depends in a more complicated way on the string coupling constants.

\subsection{General conclusions on cancelation of divergences}

In the {\it general} model there are non-vanishing divergent counterterms in all three frameworks. The occurrence of the cancelation of these divergences in the {\it fundamental string} model depends on the choice of the total energy-momentum pseudo-tensor: in the Landau-Lifshitz and canonical frameworks the divergences cancel for all $D\ge4$, while in the Dirac framework they cancel only in $D=4$. From this point of view the canonical framework seems the most convenient one, in that all divergences are proportional to the same coefficient $C$. The energy-divergence-analysis of \cite{DH,CHH} rephrases the Landau-Lihshitz framework, while the effective-action-analysis of \cite{BD12} rephrases the canonical one.

In general in all frameworks the cancelation of divergences requires actually only the conditions
\be\label{weaker}
Me^{\bt\Psi}= \La e^{\a\Psi}, \quad\quad \bt=\sqrt{\frac{2}{D-2}},
\ee
which are weaker than the defining relations \eref{fundpar} of the {\it fundamental string} model. Notice in particular that the first relation amounts to the equality between the effective string {\it tension} and {\it charge} - a property that is strictly related to the supersymmetry, more precisely $k$-symmetry, of the Green-Schwarz sigma-model action \cite{GSW}, that in absence of fermions reduces indeed to the action \eref{is}.

One has to keep in mind that, even if the divergences cancel for an appropriate choice of the coupling constants, the energy-momentum tensor must nevertheless be {\it regularized}: indeed, even in this case the {\it single} terms of the {\it bare} energy-momentum tensor are not distributions, so that it would make no sense to take their $D$-divergence. Obviously, for strings in uniform motion satisfying the conditions \eref{weaker}, the regularized field- and interaction-energy-momentum tensors {\it themselves} vanish before {\it and} after regularization (in the Landau-Lifshitz and canonical frameworks), so that for all practical purposes the regularization can be removed. However, for accelerated strings, even if the conditions \eref{weaker} hold,  the energy-momentum tensors
will be non-vanishing and the regularization must be maintained.

\section{Accelerated strings}\label{tgc}

In this section we perform a preliminary analysis of the additional problems one has   to face, when our approach is applied to accelerated strings, where its final more ambitious goal is the explicit determination of the  self-force. The general properties of the string self-forces - highly non-local functions of the whole retarded string-history - are poorly known, and in the literature one finds typically approximated explicit expressions, see for example \cite{DQ,BS,BD12bis}. For what concerns our approach, the main implication of a non-vanishing acceleration is the appearance of new divergent counterterms of the energy-momentum tensor, which in turn bear also new finite counterterms.

\subsection{New counterterms}

For strings in generic accelerated motion the velocity vectors $U_i^\m(\s)$ are no longer constant, so that their multiple covariant derivatives
\be\label{multiple}
\D_{j_1}\cdots \D_{j_p}U^\m_i
\ee
are generically non-vanishing. The main implication of this feature is that the divergent counterterms \eref{tintec} and \eref{divDc}  - to be specific from now on for simplicity we refer to the canonical framework - will receive corrections. Thanks to the manifest Lorentz- and reparameterization-invariances of our regularization, these corrections amount to additional {\it tensorial} structures in the integrands of \eref{tintec} and \eref{divDc}. Since  the divergences arise from the small-distance behavior  of the fields near the world-sheet, these new tensors must be, moreover, {\it local} expressions involving the generalized accelerations \eref{multiple}. This property restricts actually strongly the
form of these new tensors. Similarly, the non-vanishing of \eref{multiple} allows for the appearance of new {\it finite} counterterms, too.

Generically, since the indices can be contracted only with the invariant tensors $U^\m_i$ and $\eta^{\m\n}$, or their combinations,
the total number of derivatives appearing in the new tensors, acting on $\dl^D(x-y)$, or on $U^\m_i$ as in \eref{multiple}, must be {\it even}. Instead of presenting a general classification of these new structures, that would be rather cumbersome, in the following we work them out for low space-time dimensions.

{\bf\textit{D=4.}} In four dimensions we found that for a string in uniform motion the total (divergent + finite)
counterterm to be subtracted from the regularized energy-momentum tensor is (see \eref{d4can} and \eref{finitec4})
\be\label{acc4}
\wh \tau^{\m\n}\big|_{unif}=\wh\tau^{\m\n}_{int\,\ve}\big|_{div} + \wh\tau^{\m\n}_{f\ve}\big|_{div}+T^{\m\n}\big|_{fin} =
\frac{GC}{4\pi}
\int\left( \ln(\ve/L)\,l^{\m\n}+\frac{1}{2}\,\eta^{\m\n}\right)\dl^4(x-y)\sqrt{\g}\,d^2\s.
\ee
In this case the divergence is logarithmic in $\ve$, and the tensor between parenthesis is dimensionless. Consequently, since acting with derivatives lowers the length-dimension, there is no new (divergent or finite) counterterm that can show up if the string is accelerated. We conclude thus that in $D=4$ also for accelerated strings the total counterterm is given by \eref{acc4}, {\it i.e.}
$\wh \tau^{\m\n}\big|_{acc}=\wh \tau^{\m\n}\big|_{unif}$. According to \eref{tmnf}-\eref{tmnr} the total renormalized energy-momentum tensor is therefore
\be\label{tot44}
T^{\m\n}= {\cal S}'-\lim_{\ve\ra0}\left(\wh\tau_{f\ve}^{\m\n}+\wh\tau_{int\,\ve}^{\m\n} -\wh\tau^{\m\n}\big|_{acc}\right)+ \tau_{\rm kin}^{\m\n}.
\ee
According to the general strategy of section \ref{emca},
the continuity equation $\pa_\m T^{\m\n}=0$ determines then the - this time non-vanishing - self-force ${\cal S}^\m$.

{\bf\textit{D=5.}} In five dimensions for strings in uniform motion there is no finite counterterm and hence the total counterterm - adding \eref{divDc5} and \eref{tintec}  - becomes
\be\label{tot5}
\wh \tau^{\m\n}\big|_{unif}=\frac{GC}{64}
\int \frac{1}{\ve}\left(\frac{1}{2}\,\eta^{\m\n}+ \left(1-\frac{16}{\pi}\right) l^{\m\n}\right)\dl^5(x-y)\sqrt{\g}\,d^2\s.
\ee
This time the leading divergence is a simple pole $1/\ve$ and, in principle, for an accelerated string there could show up new subleading divergent and also finite counterterms, proportional to $\ln \ve$. However, for dimensional reasons the corresponding additional tensors in the integrand in \eref{tot5} would involve just {\it one} derivative, whereas, as we saw above, for covariance reasons these tensors must involve an {\it even} number of derivatives.
This means that also in $D=5$ the total counterterm for accelerated strings is given by \eref{tot5}, so that the total energy-momentum tensor is still \eref{tot44} with  $\wh \tau^{\m\n}\big|_{acc}=\wh \tau^{\m\n}\big|_{unif}$.

For generic {\it odd} dimensions $D\ge 7$, the total {\it divergent} counterterm of the uniform motion will, however, receive non-vanishing corrections if the string is accelerated, but, contrary to even $D$, there will be no {\it finite} counterterms at all. The reason is the same as in $D=5$: the finite tensors in the integrand in \eref{tot5} should have an odd number, {\it i.e.}  $D-4$, of derivatives, but an even number, {\it i.e.} two, of indices, and there are no such tensors.

{\bf\textit{D=6.}} In six dimensions for strings in uniform motion the total counterterm to be subtracted from the regularized energy-momentum tensor $\wh\tau^{\m\n}_{f\ve} +\wh\tau^{\m\n}_{int\,\ve}$ is obtained adding up \eref{divDc6}, \eref{tintec} and \eref{finc} (there is now again a finite counterterm)
\begin{align}
\wh\tau^{\m\n}\big|_{unif}=\frac{GC}{48\pi^2}
\int\bigg\{\frac{1}{\ve^2}\left(\eta^{\m\n}-5l^{\m\n}\right)& +\ln(\ve/L)\left(
\left(\eta^{\m\n}+\frac{1}{2}\,l^{\m\n}\right)\square-\pa^\m\pa^\n\right)\\
&-\left.\frac{1}{8}\,\eta^{\m\n}\,\square\right\}
\dl^6(x-y)\sqrt{\g}\,d^2\s.
\label{tot6uni}
\end{align}
As for $D=4$ and $D=5$, for accelerated strings the leading divergence - in this case $1/\ve^2$  - can not be modified, and there can be no new divergences multiplying a subleading pole $1/\ve$. This time, however, the regularized tensor $\wh\tau^{\m\n}_{f\ve}+ \wh\tau^{\m\n}_{int\,\ve}$ can produce new subleading divergences of order $\ln\ve$, multiplying two-derivative terms of the kind \eref{multiple}. For accelerated strings the total divergent counterterm has, in fact, the structure
\be\label{tot6}
\wh\tau^{\m\n}\big|_{acc}=\wh\tau^{\m\n}\big|_{unif}+
G \ln(\ve/L) \int A^{\m\n}\,
\dl^6(x-y)\sqrt{\g}\,d^2\s,
\ee
where the tensor $A^{\m\n}$, involving two derivatives, is a {\it finite} sum of terms like
\be
\begin{aligned}\label{a}
A^{\m\n}&=  a_1\, \D_iU^{\m i}\, \D_jU^{\n j}+  a_2 \,U^{\m j}\D_i\D^i U^\n_j + a_3\,  \D_iU^{\m i}\pa^\n\\
&+a_4\, \eta^{\m\n}  \Delta_iU^{\rho i}\Delta_jU_\rho^ j
+a_5 \,l^{\m\n}  \Delta_iU^{\rho i}\Delta_jU_\rho^ j +\cdots.
\end{aligned}
\ee
As the coefficient $C$ in \eref{c1},
the $a_i$ are - calculable - uniquely determined coefficients of the form \[
a_i= b_iM^2+c_i\La^2,
\]
where $b_i$ and $c_i$ are dimensionless numbers.

This procedure carries on in any dimension $D\ge6$, and the divergent part $\wh \tau^{\m\n}\big|_{acc}$ is always uniquely determined.
From these examples we see that, as the space-time dimension grows, the number  - as well as the inverse length-dimensions - of the new divergent counterterms, generalizing \eref{a}, become larger and larger and more involved. The renormalized total energy-momentum tensor \eref{tot44} is then a distribution, and the divergence of its field- and interaction-parts - for the reasons explained in section \ref{emca} - gives rise to a local expression of the kind
\begin{align}
\pa_\m\left({\cal S}'-\lim_{\ve\ra0}\left(\wh\tau_{f\ve}^{\m\n}+\wh\tau_{int\,\ve}^{\m\n} -\wh\tau^{\m\n}\big|_{acc}\right)\right)=&\nn\\
 -\int\!\big( {\cal S}^\n+{\cal S}^{\n\a_1\cdots\a_n}&\, \pa_{\a_1}\cdots\pa_{\a_n}\big) \dl^D(x-y)\sqrt{\g}\,d^2\s,
\end{align}
where we singled out the term ${\cal S}^\n$ without derivatives. The derivative terms would not give rise to a consistent  self-force, but they can be eliminated performing the finite-counterterm subtraction
\be
\wh \tau^{\m\n}\big|_{acc}\ra \wh \tau^{\m\n}\big|_{acc}
-\int {\cal S}^{\n\m\a_2\cdots\a_n}\, \pa_{\a_2}\cdots\pa_{\a_n} \dl^D(x-y)\sqrt{\g}\,d^2\s.
\ee
According to \eref{ident1} the vector ${\cal S}^\m$ identifies then the self-force.
As we saw, in $D=4$ and $D=5$ for accelerated strings no such new finite counterterms appear, but in $D=6$ the finite derivative-counterterm in \eref{tot6uni} could be modified by terms involving two derivatives, like the third term in \eref{a}. In conclusion, for any $D\ge 4$ the self-force will eventually be a multiplicative vector.

There is a last condition that ${\cal S}^\m$ must satisfy to be acceptable as a consistent self-force. The geometrical identity
 \be\label{geom}
U_{\m j}\, \Delta_i U^{\m i}=0,
\ee
stating that the acceleration is orthogonal to the velocities, requires indeed that
 \be\label{phys2}
 U_{\m j}\, {\cal S}^\m=0,
 \ee
as a consistency condition for the string equation \eref{eqself}.
Though necessary for the existence of an internally consistent string dynamics compatible with energy-momentum conservation, the condition \eref{phys2} is not guaranteed {\it a priori}\footnote{ Notice that, thanks to reparameterization invariance, the infinite bare self-force \eref{ill} would satisfy \eref{phys2} automatically.}. It could very well be that in order to satisfy \eref{phys2} -  in line with our strategy that fixes the energy-momentum tensor only in the complement of the world-sheet - one must subtract further finite counterterms of the form \eref{a}. We will come back to this point in the concluding section.

\subsection{Uniqueness: finite counterterms}\label{ufc}

Eventually we address the problem of whether the dynamics of self-interacting strings,  codified  by the self-force, derived according to the above procedure is uniquely determined. Within our approach this question is tied intimately to the uniqueness properties of the renormalized total energy-momentum tensor  $T^{\m\n}$ \eref{tot44}, that is actually subject to an {\it ultimate} finite-counterterm-ambiguity, {\it i.e.} the freedom of modifying it according to
\be\label{modi}
 T'^{\m\n} = T^{\m\n}+  G
\int  I^{\m\n}\,\dl^D(x-y)\sqrt{\g}\,d^2\s,
\ee
where, for a generic $D\ge6$, the tensor $I^{\m\n}$ is made out of the higher dimensional kinematical analogues of \eref{a}. To say it again, the freedom of adding such terms arises from the fact that in the divergent counterterm \eref{tot6} the tensors \eref{a} appear multiplied by the finite coefficients $\ln L$, required for dimensional reasons, which are actually {\it arbitrary}.

The tensor $I^{\m\n}$ is, however, constrained by two consistency conditions. In the first place the divergence of the added term must have the structure
\be\label{imn}
\pa_\m\int  I^{\m\n}\dl^D(x-y)\sqrt{\g}\,d^2\s=
\int  I^{\m\n}\,\pa_\m\dl^D(x-y)\sqrt{\g}\,d^2\s =- \int  I^\n\, \dl^D(x-y)\sqrt{\g}\,d^2\s,
\ee
for some {\it multiplicative} vector $I^\n$. Only in this case the modification \eref{modi} can indeed give rise, through $\pa_\m T'^{\m\n}=0 $, to the modified self-force (see again \eref{ident1})
\[{\cal S}'^\m=  {\cal S}^\m+ G I^\m,
\]
and hence to a physically inequivalent dynamics. The second condition is that the vector $I^\m$ must respect the geometrical identity \eref{phys2}, {\it i.e.}
\be\label{geom2}
 U_{\m i}I^\m=0.
\ee

There are of course a lot of tensors $I^{\m\n}$ satisfying \eref{imn} and \eref{geom2} trivially, namely derivative operators of the kind  $I^{\m\n}= W^{\rho\m\n}\pa_\rho$, with $W^{\rho\m\n}=- W^{\m\rho\n}$. In these cases \eref{imn} would hold simply with $I^\m=0$, so that the self-force would remain unaltered. For a non-vanishing $I^\m$ the properties \eref{imn} and \eref{geom2} turn out to be very restrictive. For $D=4$, for example, there would be only the trivial choice $I^{\m\n}= bM^2l^{\m\n}$ - with $b$ a dimensionless constant - that corresponds merely to a redefinition of the string tension\footnote{The choice between $M$ and $\La$  in $I^{\m\n}$ is purely conventional, since they have both the dimension of an inverse length squared.}. No such tensor exists for $D=5$. For the simplest non-trivial dimension $D=6$  the tensor $I^{\m\n}$ must be a combination of terms like \eref{a}. We found actually only {\it one} tensor  satisfying  \eref{imn} and \eref{geom2}, whose construction is as follows.

Consider the functional of the string coordinates $y^\m(\s)$
\be\label{acc2}
L[y]=bM^2 \int \D_iU^{\m i}\D_jU^{\n j}\,\eta_{\m\n}\sqrt{\g}\,d^2\s,
\ee
which can be seen to be the unique reparameterization- and Lorentz-invariant local functional containing {\it two} derivatives ($b$ is a dimensionless constant as above). Construct its {\it curved} counterpart $L_g[y]$, obtained from \eref{acc2} replacing everywhere the flat matric $\eta_{\m\n}$ with $g_{\m\n}$, {\it e.g.} $\g_{ij}\ra \G_{ij}$ etc. Define then the tensor $I^{\m\n}$ through the functional derivative
\be\label{func}
-\frac{2}{\sqrt{g}}\,\frac{\dl L_g[y]}{\dl g_{\m\n}(x)}\,\bigg|_{g=\eta}=  \int I^{\m\n}\,\dl^6(x-y)\sqrt{\g}\,d^2\s,
\ee
and introduce the world-sheet vector
\be\label{wv}
I_\m=-\frac{\dl L[y]}{\dl y^\m}.
\ee
Then from $D=6$ diffeomorphism invariance of $L_g[y]$ it follows that the so defined tensors $I^{\m\n}$  and $I^\m$ satisfy \eref{imn}, and from the world-sheet reparameterization invariance of $L[y]$ it follows  that $I^\m$ satisfies \eref{geom2}.
Given the structure of \eref{acc2}, the tensor $I^{\m\n}$ defined in \eref{func} has precisely the form \eref{a}, containing in particular two derivatives.

In conclusion, in $D=6$ after renormalization the effective string dynamics is determined modulo the self-force $GI^\m$, defined via \eref{acc2} and \eref{wv}, which introduces thus a {\it new  coupling constant} in the theory, namely the coefficient $b$.  It is clear that this new interaction amounts to the the replacement of the string kinetic action
\be\label{kinplus}
-Me^{\bt \Psi} \!\int\! \sqrt{\g}\,d^2\s\quad \ra \quad-Me^{\bt \Psi}\! \int\!
\sqrt{\g}\,d^2\s +G\,L[y],
\ee
which introduces new local higher-derivative self-interactions.

We stress that the occurrence of this new interaction is an unavoidable consequence of the renormalization process itself: it is the simple mathematical statement that a divergent term is intrinsically defined modulo finite terms. While in the {\it generic} string model the coupling constant $b$ appears to be arbitrary, it may happen that in the {\it fundamental string} model - as a classical version of superstring theory - this coupling constant must vanish, or be fixed to some  specific value.
For higher {\it even} dimensions $D\ge8$, the invariant self-interaction functionals of the type \eref{acc2} contain a growing number of derivatives, and so the number of independent functionals, and thus new coupling constants, grows rapidly.

\subsection{Uniqueness: framework-dependence}

A further source of non-uniqueness of the self-force could arise from the dependence of the whole procedure on the choice of the {\it framework}, that is, the freedom in the choice of the total energy-momentum pseudo-tensor. Before regularization each pair of the three total energy-momentum tensors \eref{tott}, \eref{ttilde}, \eref{cano1} - {\it i.e.} $\tau^{\m\n}$, $\w\tau^{\m\n}$ and $\wh\tau^{\m\n}$ - are tied by a relation of the kind (see \eref{t12} and  \eref{cano})
\be\label{12}
\tau_{(2)}^{\m\n}-\tau_{(1)}^{\m\n}=\pa_\rho Z^{\rho\m\n}+\mbox{ field equations of motion},
\ee
where  $Z^{\rho\m\n}=-Z^{\m\rho\n}$. Since $\tau_{(2)}^{\m\n}$ and $\tau_{(1)}^{\m\n}$ contain as kinetic part of the string the same tensor $\tau_{\rm kin}^{\m\n}$ \eref{tslin}, the equation \eref{ident1} would give rise  {\it formally} to the same self-force, since $Z^{\rho\m\n}$ simply drops out from that equation.

At the regularized level there is, however, a subtlety that may arise. If we consider the regularized energy-momentum tensors $\tau_{(i)\ve}^{\m\n}$, constructed with the regularized fields $f_\ve$ \eref{fxe}, the relation \eref{12} would still hold for the $\tau_{(i)\ve}^{\m\n}$, where the regularized tensor $Z^{\rho\m\n}_\ve$ is still antisymmetric in its first two indices. Consequently, since {\it distributional} derivatives always commute,  the identity
$\pa_\m\pa_\rho Z^{\rho\m\n}_\ve=0$ holds true also for finite $\ve$. But this time the field equations of motion at the r.h.s of \eref{12} would amount to {\it regularized} equations of motion  - not identically vanishing -  which as $\ve\ra 0$ could give rise in \eref{12} to (divergent and finite) contributions supported on the world-sheet.  Clearly - by construction - the divergent contributions are removed by our renormalization procedure, but there could remain finite parts which in \eref{ident1} could give rise, in turn, to different self-forces. We believe actually that such discrepancies do not arise, in that we conjecture that the unique freedom of the dynamics of self-interacting strings is represented by the {\it universal} local self-coupling \eref{acc2}, arising directly from the renormalization process. After all, a part from this self-coupling, there should exist a unique well-defined dynamics ``associated'' to the linearized {\it formal} action \eref{fands}, as it happens {\it e.g.} for self-interacting charged particles in $D=4$ \cite{R}. Probably
a definitive test of this conjecture can be provided only through a direct calculation of the self-forces in the three frameworks.

\section{Conclusions}\label{cr}

Due to the presence of ultraviolet divergences, the derivation of the dynamics of self-interacting strings, taking back-reaction into account, can not be founded on a variational principle - based on a {\it canonical, local} and {\it finite} action. In absence of such a principle, in this paper we proposed a universal procedure for the derivation of this dynamics in arbitrary dimensions that, $i)$ incorporates by construction energy-momentum conservation and, $ii)$ gives rise automatically to a finite self-force.

We tested three versions - frameworks - of this procedure, in the pilot program of flat strings, where its main characteristics and advantages emerge clearly: manifest reparameterization and Lorentz-invariance, separability and locality of divergences, the presence of subleading divergences, the need of local finite counterterms to derive a multiplicative self-force. All frameworks give in this case rise to a consistent total conserved and covariant energy-momentum tensor and to a vanishing self-force. In the {\it fundamental string} model for all $D\ge4$ we retrieved in the Landau-Lifshitz and canonical frameworks the cancelation of all ultraviolet divergences, including also the entire set of subleading divergences - cancelations that incorporate the tension non-renormalization retrieved in previous approaches and foreseen by superstring theory. In the Dirac framework this cancelation occurs only in $D=4$. The cancelation of ultraviolet divergences is thus in general framework-dependent; in particular the failure of the non-renormalization of the string tension in the Dirac framework for $D>4$ may signal that this framework is unable to furnish the correct classical counterpart of superstring theory. This feature is however not a problem for what concerns the construction of a correct {\it classical} theory, since in our approach a consistent dynamics of self-interacting strings can be derived independently of the values of the coupling constants.

We faced the problem of the derivation of the self-force ${\cal S}^\m$ for accelerated strings, analyzing in particular the form of the new divergent counterterms. We found that all derivative-self-forces can be eliminated through the subtraction of finite counterterms from the energy momentum tensor.  The explicit check whether or not the resulting multiplicative self-forces satisfy the orthogonality condition \eref{phys2}, is tied to the future program of determining the self-forces explicitly. With this respect the simplest case is $D=4$, since there the exact total counterterm \eref{acc4} is known and receives no acceleration-induced corrections. The validity of the conjectured relation \eref{phys2} is based on a physical credo: would it not hold, there would exist no dynamics of self-interacting strings compatible with energy-momentum conservation.
From this point of view the situation is the same as for a charged self-interacting particle in $D=4$: {\it a priori} there is no reason that the Lorentz-Dirac self-force
\[
{\cal S}^\m=\frac{e^2}{6\pi}\,\bigg(\frac{d^2u^\m}{ds^2}+ \left(\frac{du}{ds}\right)^2 u^\m\bigg),
\]
derived from the requirement of conservation of the renormalized total energy-momentum tensor \cite{R,LM}, is orthogonal to the four-velocity $u^\m$, but eventually it turns out to be so.

Regarding the uniqueness of our construction we revealed the appearance of a finite number of new {\it local} self-interactions - none of them occurs in $D=4$ and $D=5$ and only a single self-interaction term occurs in $D=6$ - that are tied intrinsically to the renormalization process. Correspondingly we believe that this is the unique source of ambiguity in the dynamics of self-interacting strings, so that the self-force is not framework-dependent. Again, this statement requires a test through an explicit computation.

\paragraph{Acknowledgments.}

This work is supported in part by the {\it INFN Iniziativa Specifica STEFI}.

\vskip0.5truecm




\end{document}